\documentclass[a4paper,12pt]{article}
\usepackage[T1,T2A]{fontenc}
\usepackage[utf8]{inputenc}
\usepackage[ukrainian,english]{babel}

\usepackage[a4paper]{geometry}
\usepackage{amsmath}

\usepackage{graphicx}
\usepackage[english]{babel}
\usepackage{authblk}

\title{\uppercase{On fast charged particle scattering by periodic atomic planes: quadratic potential corrections}}

\usepackage{lipsum}

\usepackage{multicol}
\usepackage{subcaption}
\usepackage{amssymb}
\usepackage{placeins}

\author[1]{V.D.~Omelchenko \thanks{koriukina@kipt.kharkov.ua; ORCID: 0000-0002-5113-3028 }}
\affil[1]{National Science Center ``Kharkiv Institute of Physics and Technology'', Kharkiv, Ukraine}

\begin{document}
\maketitle

\begin{abstract}
In this paper, the approach for considering fast charged particles scattering on targets of complex structure, which contains some isolated substructures, was expanded to account quadratic potential terms. Based on this approach, the differential cross section for scattering on the set of parallel planes with uniformly distributed atoms in each plane was obtained. It was shown that for this case the differential scattering cross section splits into coherent and incohent cross sections in the eikonal approximation analogously with the Born approximation.

\end{abstract}

\section{Introduction}

Among key papers related to fast charged particles scattering one may highlight the following papers and ideas. In the work \cite{Bragg12_1}, Bragg showed dependence of wave diffraction patterns on the structure of the crystal which the wave passes. Ter-Mikaelian, in his turn, demonstrated in \cite{MLT72} that there exists the diffraction of high-energy particles, represented as plane waves, on crystal targets. The key idea of Ter-Mikaelian's work was in showing that despite de Broil wave length of fast particles being of other order of value compared to characteristic distances between atoms in crystal, the diffraction of fast charged particles represented as plane waves is still observable. This idea indicates the possibility of quantum effects manifestation in processes involving high-energy particles. The calculations in \cite{MLT72} were performed in the Born approximation of quantum electrodynamics resulting in differential scattering cross section which is easy enough to obtain and analyze but having relatively narrow applicability region. 

The eikonal approximation \cite{AIA96} compared with Born approximation has wider range of application but for real potential of targets the calculations become impossible to be performed analytically. The way to handle this difficulty was suggested by Glauber \cite{Glauber}. It consisted in expansion of the differential scattering cross section in the series of potential-related functions and making the necessary averaging of the series. By necessary averaging we mean the need to average the differential cross section over atoms positions in the target so that the structure of the target was involved in calculations in simplified way. Also it is worth mentioning Lindhard's idea of continuous potential \cite{Lind} that suggested replacing the complicated potential of the target with its averaged potential. This approach significantly simplifies the calculation but does not account for all the effects. In works of Akhiezer and Shul'ga (see \cite{AIA96} and the references there) there were presented the considerations on the eikonal approximation usage for scattering problems. One of main ideas there was the implementation of limiting case of the eikonal approximation which is quasi-classical approach enabling to consider relatively complicated potentials in simpler ways.

Also, the works on the rainbow scattering are of particular interest since on the one hand this type of scattering appears to be sensitive to the structure of the target  \cite{Petrovic04}, and on the other hand the quantum rainbow scattering is potentially observable quantum effect. In the papers \cite{FominR, Petrovic13}, Fomin, Petrović and other authors considered manifestation of quantum rainbow scattering in interaction with atomic string and nanotubes.

Ideas of using the eikonal approximation and quasi-classical approach were also used in Shul'ga's works. In the mentioned papers, there were obtained the differential scattering cross section for set of parallel atomic planes in the model of uniform distribution of atoms in planes in the Born approximation \cite{Born}; the differential scattering cross section on a single atomic plane in the eikonal approximation and limiting transition to quasi-classical approximation for this case \cite{Eik} and the rainbow scattering on a single plane in the eikonal approximation \cite{RainC}. This works were continued in the recent paper \cite{PerPl1}, where using the continuous potential idea, the expression for the differential scattering cross section was significantly simplified. In the present paper, we consider the quadratic potential correction for the result obtained in \cite{PerPl1}. Considering quadratic on potential terms leads us to including incoherent effects to the differential scattering cross section.

\section{Problem formulation}
\begin{figure}[h]
      \centering
	   \includegraphics[width=0.5\textwidth]{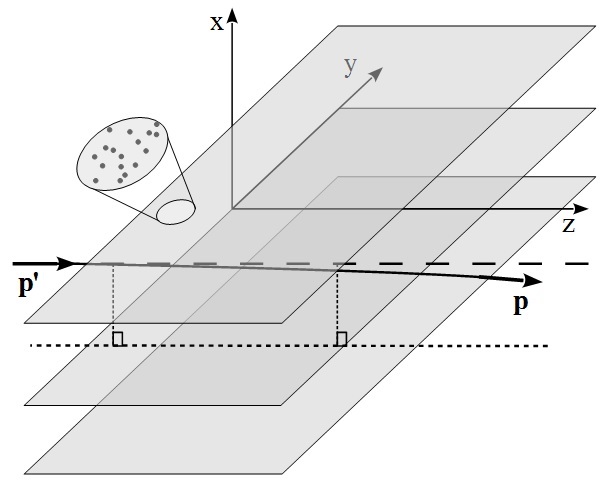}
	\caption{The fast charged particle scattering on the set of parallel planes with uniformly distributed atoms in each plane}
 \label{fig_np}
\end{figure}

Let us consider a fast charged particle with initial momentum $\vec{p}$ parallel to $z$-axis, incident on a target. The target is a set of parallel planes with uniform distribution of atoms within each of them (Fig. \ref{fig_np}). This problem is the model for particles scattering while moving along atomic planes in crystals when the influence of separate atomic strings could be neglected. Let the target consist of $N$ atoms: $N=N_x N_p$, where $N_x$ is the number of atomic planes in the target, $N_p$ is the number of atoms in each plane (assuming each plane has the same amount of atoms). Let the planes be equidistant with spacing $a$. The size of planes is given by $L_y$, $L_z$ along $y$, $z$ axes respectively. The number of atoms in each plane is related to their sizes through 2D-concentration of atoms in the plane $n_{yz}$:
\begin{eqnarray}\label{eq0}
N_p=n_{yz} L_y L_z.
\end{eqnarray}

In the eikonal approximation, the distribution of atoms along $z$-axis does not influence the differential scattering cross section, so we will not specify it. The probability of each atom in the target to poses a certain position along $y$-axis, due to the uniform distribution of atoms within planes, is
\begin{eqnarray}\label{eq0_1}
g(y_{0})= \begin{cases}\frac{1}{L_y}, \ -\frac{L_y}{2}\leq y_0 \leq \frac{L_y}{2};\\
0, \ |y_0|>\frac{L_y}{2}.
\end{cases}
\end{eqnarray}
Let us consider the case of a crystal without thermal vibrations for simplicity.

\section{Differential cross section of fast charged particles scattering on set of parallel atomic planes}

Let us consider scattering of fast charged particles on the target of $N$ atoms which has the potential $U^{(N)}(\vec{r})$  
\begin{eqnarray}\label{eq1}
U^{(N)}(\vec{r})=\sum_{n=1}^N u(\vec{r}-\vec{r}_n),\end{eqnarray}
where $u(\vec{r}-\vec{r}_n)$ is the potential of individual atom in the target with position $\vec{r}_n$, $\vec{r}=(x,y,z)$. 

Let us assume the potential of individual atom to be screened Coulomb potential
\begin{eqnarray}\label{eq1_1_1}
u(\vec{r})=\frac{Ze}{r}\exp \left(-\frac{r}{R} \right),
\end{eqnarray}
where $R$ is the screening radius, $Ze$ is the charge of atom nucleus.

For considering the scattering problem we need to know the wave function of a scattered particle, which is proportional to $\exp[i\tilde{\chi}_0^{(N)}(\vec{\rho})]$ in the eikonal approximation, where 
\begin{eqnarray}\label{eq1_1}
\tilde{\chi}^{(N)}_0(\vec{\rho})=-e\int_{-\infty}^{\infty} dz U^{(N)}(\vec{\rho},z)\end{eqnarray}and $\vec{\rho}=(x,y)$. 

In this case, the differential scattering cross section \cite{AIA96} 
\begin{eqnarray}\label{eq1_2}
\frac{d\sigma}{d^2q_{\perp}}= |{a}(\vec{q}_{\perp})|^2,\end{eqnarray}
can be expressed through the scattering amplitude
\begin{eqnarray}\label{eq2} 
a(\vec{q}_{\perp})=\frac{i}{2\pi} \int_{-\infty}^{\infty} dx \int_{-\infty}^{\infty} dy e^{i \vec{q}_{\perp} \vec{\rho}} \left\{ 1- \exp\left[ i \tilde{\chi}^{(N)}_0 (\vec{\rho}) \right] \right\}.
\end{eqnarray} 
where $\vec{q}_{\perp}=(q_x,q_y)$ is the momentum transfer along the corresponding axes. 

Let us define the $\tilde{\chi}_0^{(1)}$-function for an individual atom in the target as 
\begin{eqnarray}\label{eq3}
\tilde{\chi}_0^{(1)} (\vec{\rho}-\vec{\rho}_n)=-e\int_{-\infty}^{\infty} dz u(\vec{\rho}-\vec{\rho}_n,z).\end{eqnarray}
The $\tilde{\chi}_0^{(1)}$-function does not depend on $z_n$ due to substitution $z-z_n \rightarrow z$ which does not change the limits of integration equal to $\pm \infty$. This property of the integral leads to independence of the scattering amplitude on $z$-component of atoms positions in the target. From \eqref{eq1}, \eqref{eq1_1} and \eqref{eq3} we can derive
\begin{eqnarray}\label{eq4}
\tilde{\chi}^{(N)}_0 (\vec{\rho})=\sum_{n=1}^N \tilde{\chi}_0^{(1)} (\vec{\rho}-\vec{\rho}_n).
\end{eqnarray}
Using \eqref{eq2}, the expression for differential scattering cross section \eqref{eq1_2} can be written as  
\begin{eqnarray}\label{eq4_1}
\frac{d\sigma}{d^2q_{\perp}}=\frac{1}{4\pi^2} \int_{\mathbb{R}^4}d^2 \rho d^2\rho' e^{i\vec{q}_{\perp}(\vec{\rho}-\vec{\rho}')} \left(1-e^{i\tilde{\chi}^{(N)}_0(\vec{\rho})} -e^{-i\tilde{\chi}^{(N)}_0(\vec{\rho}')}+ e^{i[\tilde{\chi}^{(N)}_0(\vec{\rho})-\tilde{\chi}^{(N)}_0(\vec{\rho}')]} \right),
\end{eqnarray}
where $\int_{\mathbb{R}^4}d^2 \rho d^2\rho'...$ denotes the integral over the entire real four-dimensional volume.

Considering $\tilde{\chi}^{(N)}_0$-function has a very complicated form, we need to average the expression for differential scattering cross section \eqref{eq4_1} over atoms positions in the target. So, we redefine the differential scattering cross section in the following way 
\begin{eqnarray}\label{eq4_2}
\frac{d\sigma}{d^2q_{\perp}}=\frac{1}{4\pi^2} \int_{\mathbb{R}^4}d^2 \rho d^2\rho' e^{i\vec{q}_{\perp}(\vec{\rho}-\vec{\rho}')} \langle 1-e^{i\chi^{(N)}_0(\vec{\rho})} -e^{-i\chi^{(N)}_0(\vec{\rho}')}+ e^{i[\chi^{(N)}_0(\vec{\rho})-\chi^{(N)}_0(\vec{\rho}')]} \rangle,
\end{eqnarray}
where  the brackets $\langle ... \rangle$ denote averaging, e.g. $\langle ... \rangle=\int_{-\infty}^{\infty}dy_0 g(y_0) ...$.

Using Glauber's idea \cite{Glauber} in our previous paper \cite{PerPl1}, we showed that for scattering on the set of parallel atomic planes
\begin{eqnarray}\label{eq6}
\langle e^{i[\chi^{(N)}_0(\vec{\rho})-\chi^{(N)}_0(\vec{\rho}')]} \rangle =  \exp \left\{N_p \left[i \sum_{j=1}^{N_x} \langle \chi_{(j)}-\chi'_{(j)} \rangle - \right. \right.  \nonumber \\  
\left. \left. -\frac{1}{2} \sum_{j=1}^{N_x} \langle \left( \chi_{(j)}-\chi'_{(j)} \right)^2 \rangle + \frac{1}{2} \sum_{j=1}^{N_x}\langle \chi_{(j)}-\chi'_{(j)} \rangle ^2 + ... \right]\right\},
\end{eqnarray}
where $\chi_{(j)}=\chi_0^{(1)}(x-aj,y-y_0)$, $\chi'_{(j)}=\chi_0^{(1)}(x'-aj,y'-y_0)$ are $\chi_0$-functions for a single plane. In this definitions $\langle \chi_0^{(1)}(\vec{\rho}) \rangle =N_p \langle \tilde{\chi}_0^{(1)}(\vec{\rho}) \rangle$. 
Analogously, 
\begin{eqnarray}\label{eq6}
\langle e^{i\chi^{(N)}_0(\vec{\rho})} \rangle =  \exp \left\{N_p \left[i \sum_{j=1}^{N_x} \langle \chi_{(j)} \rangle -\frac{1}{2} \sum_{j=1}^{N_x} \langle  \chi_{(j)}^2 \rangle + \frac{1}{2} \sum_{j=1}^{N_x}\langle \chi_{(j)} \rangle ^2 + ... \right]\right\}.
\end{eqnarray}

Considering only linear on potential terms in \eqref{eq4_2} for parallel atomic planes distanced from each other at $a \gg R$ leads us to the following expression \cite{PerPl1}
\begin{eqnarray}\label{eq7}
\frac{d\sigma^{(N_x)}_{eik}}{dq_{x}}=\frac{d\sigma^{(1)}_{eik}}{dq_{x}} D_{N_x},
\end{eqnarray}
where $D_{N_x}=\Big| \sum_{k=1}^{N_x} e^{i q_{x}x_k} \Big|^2$, $\frac{d\sigma^{(1)}_{eik}}{dq_{x}}$ is the differential cross section of fast charged particles scattering on a single atomic plane located at $x=0$ and centered symmetrically along $y$-axis.  
Formula \eqref{eq7} showed that for planes which could be considered isolated ($a \gg R$) the differential scattering cross section can be written as a product of differential cross section of scattering on a single atomic plane and the structure factor $D_{N_x}$. Considering cross section in this form significantly simplified the calculations agreeing well with numerical integration of  \eqref{eq4_2} (with accuracy up to linear potential terms). Considering only linear potential terms in \eqref{eq7} is equivalent to considering the scattering problem in the continuous potential approximation which corresponds to describing coherent scattering. Quadratic potential terms responsible for the cross section ``smearing’’ (as shown in \cite{Eik}) are connected with incoherent scattering. Considering only linear terms allowed us to find general patterns and features of the scattering process. But including quadratic terms will allow us to obtained more precise and more relevant for comparison with experiments expression for differential scattering cross section.

We notice that $\frac{1}{2} \sum_{j=1}^{N_x}\langle \chi_{(j)}-\chi'_{(j)} \rangle ^2$ is proportional to $1/L_y^2$ which makes it relatively small compared to  $\frac{1}{2} \sum_{j=1}^{N_x} \langle \left( \chi_{(j)}-\chi'_{(j)} \right)^2 \rangle \sim 1/L_y$ , so we can neglect the term $\frac{1}{2} \sum_{j=1}^{N_x}\langle \chi_{(j)}-\chi'_{(j)} \rangle ^2$.

Let us consider \eqref{eq4_2} in the following form
\begin{eqnarray}\label{eq8}
\frac{d\sigma}{d^2q_{\perp}}=\frac{1}{4\pi^2} \int_{\mathbb{R}^4}d^2 \rho d^2\rho' \ e^{i\vec{q}_{\perp}(\vec{\rho}-\vec{\rho}')} \left(1-e^F-e^{F'}+e^F e^{F'}e^G \right),
\end{eqnarray}
where $F(x)=i \sum_{j=1}^{N_x} \langle \chi_{(j)}\rangle -\frac{1}{2} \sum_{j=1}^{N_x} \langle  \chi_{(j)}^2 \rangle$, $F'(x')=F^*(x')$, $G(x,x',\Delta y)=2 \sum_{j=1}^{N_x} \langle  \chi_{(j)} \chi'_{(j)} \rangle$. Later, calculating mentioned here functions, we will show that $\langle  \chi_{(j)} \rangle$, $\langle  \chi_{(j)}^2 \rangle$ depend only on $x$ and $\langle  \chi_{(j)} \chi'_{(j)} \rangle$ depends on $x$, $x'$ and only on the difference $y-y'=\Delta y$ for the model of uniform distribution of atoms in each plane of the target.
Expression $\left(1-e^F-e^{F'}+e^F e^{F'}e^G \right)$  can be rewritten equivalently as
\begin{eqnarray}\label{eq9}
1-e^F-e^{F'}+e^F e^{F'}e^G = \left(1-e^F\right)\left(1-e^{F'} \right)-\left(1-e^F \right)\left(1-e^{F'} \right)\left(1-e^G \right)+ \nonumber \\ + \left[\left(1-e^F \right)\left(1-e^G \right) +\left(1-e^{F'} \right)\left(1-e^G \right) \right]-\left(1-e^G \right).
\end{eqnarray}
Due to $F'(x')=F^*(x')$, the expression in the square brackets in \eqref{eq9} is 
\begin{eqnarray}\label{eq10}
\int_{\mathbb{R}^4}d^2 \rho d^2\rho' \ e^{i\vec{q}_{\perp}(\vec{\rho}-\vec{\rho}')} \left[\left(1-e^F \right)\left(1-e^G \right) +\left(1-e^{F'} \right)\left(1-e^G \right) \right]= \nonumber \\
= 2 Re \int_{\mathbb{R}^4}d^2 \rho d^2\rho' \ e^{i\vec{q}_{\perp}(\vec{\rho}-\vec{\rho}')} \left(1-e^F \right)\left(1-e^G \right).
\end{eqnarray}

To make the differential scattering cross section comparable with result in \cite{PerPl1}, let us integrate the differential cross section \eqref{eq8} over $q_y$. Let us consider the part of \eqref{eq8} depending on $q_y$, $y$, $y'$: 
\begin{eqnarray}\label{eq11}
\int_{-\infty}^{\infty}dq_y\int_{\mathbb{R}^2}dy dy' \ e^{i{q}_{y}(y-y')}e^G=2 \pi L_y \int_{-\infty}^{\infty}d \Delta y \ \delta\left(\Delta y \right) e^{G(x,x',\Delta y)}.
\end{eqnarray}
Expression \eqref{eq11} was calculated with the idea that integral $\int_{-\infty}^{\infty}dy$ makes sense only in range $y \in [-L_y/2,L_y/2]$. This happens due to the fact that screened potential of atoms in the target has a short range of influence, so it does not influence incident particles which pass far from the target. So, this integral should be considered as if $\int_{-\infty}^{\infty}dy=L_y$. After integrating \eqref{eq11} over $\Delta y$ we obtain   
\begin{eqnarray}\label{eq12}
\int_{-\infty}^{\infty}dq_y\int_{\mathbb{R}^2}dy dy' \ e^{i{q}_{y}(y-y')}e^G=2 \pi L_y  e^{\tilde{G}}.
\end{eqnarray}
where $\tilde{G}(x,x')=G(x,x',\Delta y=0)$. 
So, after integration over $q_y$, \eqref{eq8} becomes
\begin{eqnarray}\label{eq12_1}
\frac{d\sigma}{dq_{x}}=\frac{L_y}{2\pi} \int_{\mathbb{R}^2}dx dx' \ e^{i {q}_{x}(x-x')} \left(1-e^F-e^{F'}+e^F e^{F'}e^{\tilde{G}} \right),
\end{eqnarray}
which is equivalent to 
\begin{eqnarray}\label{eq13}
\frac{d\sigma}{dq_{x}}=\frac{L_y}{2\pi} \left\{ I_0 -I_1 - I_2 +I_3  \right\},
\end{eqnarray}
where $\{I_j \}_{j=0}^3$ denotes following integrals: $I_0=\int_{\mathbb{R}^2}dx dx' \ e^{i{q}_{x}(x-x')} \left(1-e^F\right)\left(1-e^{F'} \right)$, \linebreak $I_1=\int_{\mathbb{R}^2}dx dx' \ e^{i{q}_{x}(x-x')} \left(1-e^F \right)\left(1-e^{F'} \right)\left(1-e^{\tilde{G}} \right)$, $I_2=\int_{\mathbb{R}^2}dx dx' \ e^{i{q}_{x}(x-x')} \left(1-e^{\tilde{G}} \right)$, \linebreak $I_3=2Re\int_{\mathbb{R}^2}dx dx' \ e^{i{q}_{x}(x-x')} \left(1-e^F \right)\left(1-e^{\tilde{G}} \right)$. We note that $I_0$ may be calculated in easier way as $I_0=\Big|\int_{-\infty}^{\infty}dx \ e^{i{q}_{x}x} \left(1-e^F\right) \Big|^2$ analogously to the expression obtained in \cite{PerPl1}. Though, $I_0$ here is defined with higher precision and generally speaking does not coincide with the one in \cite{PerPl1}.

\section{Integrals connected with the problem}
\sloppy
Approach used in the article \cite{PerPl1} can be genelarized as follows. Suppose that we consider a target consisting of isolated structures (atoms, atomic planes, strings, etc.). A structure in the target can be considered isolated if the distance from this structure to other structures is much larger than the screening radius of the atoms that constitute this structure. Let us consider the function $H\left(\sum_m f_m(X) \right)$, where $f_m(X)$ is a potential-dependent function corresponding to the $m$-th isolated structure in the target; $X$ denotes the spatial variables on which this function depends. If $f_m$ is arranged so that it has much larger values in coordinates close to the $m$-th isolated structure than in coordinates distanced from it by approximately half the distance between such structures, then we can approximately assume that at each coordinate point only one of the structures makes a significant contribution to the value of the function $H$:
\begin{eqnarray}\label{eq13_1}
H\left(\sum_m f_m \right) \approx H\left(f_m \right) \Big|_{X \in Y_m},
\end{eqnarray}
where $Y_m$ denotes the coordinate region around $m$-th isolated structure and is chosen so that $f_m$ has much smaller values at its edges and outside this region, compared to the values inside the region. Also, the position of an individual structure $X_m$ in $X$ coordinates: $X_m \in Y_m$. The set of all $\left\{Y_m \right\}$ is chosen so that regions with different $m$ do not intersect (except at the boundaries). 

We also impose a condition on the function $H$: let the function $H\left(f_m \right)$ inside each region $Y_m$ take much larger values than when approaching the edges. When all the specified conditions are met, we can approximately assume that 
\begin{eqnarray}\label{eq13_2}
\int_{\mathbb{R}^{dimX}} dX \ e^{iQ X} H\left(\sum_m f_m(X)\right) \approx \sum_m \int_{Y_m} dX \ e^{iQ X} H(f_m(X)),
\end{eqnarray}
where $Q X$ denotes the scalar product of the set of variables $X$ with a certain parameter $Q$, 
$\dim X$ is the dimension of the set $X$, which equals the number of variables on which the functions $f_m$ depend, 
and $\mathbb{R}^{\dim X}$ denotes the entire real volume of that dimension, meaning that each variable in $X$ takes all values from $-\infty$ to $+\infty$.

Equation~\eqref{eq13_2} indicates that the integral over the entire real volume $\mathbb{R}^{\dim X}$ can be decomposed into a sum of integrals over the regions $\{Y_m\}$, where the functions $\{f_m\}$ are significantly different from zero. This is due to the fact that $f_m \approx 0$ outside the boundaries of $X_m$. Condition that each $H\left(f_m \right)$ rapidly decays approaching edges of $Y_m$ leads to
\begin{eqnarray}\label{eq13_3}
\int_{Y_m} dX \ e^{iQ X} H(f_m(X)) \approx \int_{\mathbb{R}^{dimX}} dX \ e^{iQ X} H(f_m(X)).
\end{eqnarray}
Performing the substitution  $\tilde{X} = X-X_m$ results in
\begin{eqnarray}\label{eq13_4}
\int_{\mathbb{R}^{dimX}} dX \ e^{iQ X} H(f_m(X)) = e^{iQX_m} \int_{\mathbb{R}^{dimX}} dX \ e^{iQ \tilde{X}} H(f_0(\tilde{X})).
\end{eqnarray}
In \eqref{eq13_4}, $f_0(\tilde{X})$ corresponds to the function $f_m$ for an isolated structure located at the coordinate origin with respect to all variables in the set $X$. From \eqref{eq13_2} and \eqref{eq13_4} it follows that
\begin{eqnarray}\label{eq13_5}
\int_{\mathbb{R}^{dimX}} dX \ e^{iQ X} H\left(\sum_m f_m(X)\right) \approx \left(\sum_m e^{iQX_m} \ \tilde{1}_{H(Y_m)} \right) \left(\int_{\mathbb{R}^{dimX}} dX \ e^{iQ \tilde{X}} H(f_0(\tilde{X})) \right),
\end{eqnarray}
where $\tilde{1}_{H(Y_m)}$ is an analogue of an indicator function, which takes the value $0$ when we assume $H(Y_m) \approx 0$, and $1$ otherwise. This function “cuts off’’ the contributions of the regions where $H(Y_m) \approx 0$.

On the right-hand side of \eqref{eq13_5}, both structures inside the parentheses are evaluated independently. Thus, taking an integral of the form $\int_{\mathbb{R}^{dimX}} dX \ e^{iQ X} H\left(\sum_m f_m(X)\right)$ is significantly simplified.

Let us also consider a generalization where
\begin{eqnarray}\label{eq13_5_1}
H\left(\sum_m f_{(1),m}(X), \sum_m f_{(2),m}(X), ..., \sum_m f_{(M),m}(X) \right) \equiv H\left(\{\sum_m f_{(s),m}(X)\}_{s=1}^M \right), 
\end{eqnarray}
where $\{f_{(s),m}(X)\}_{s=1}^M$ is a set of potential-dependent functions corresponding to the $m$-th isolated structure in the target. Let us assume that all $f_{(s),m}$ are arranged analogously to $f_m$ in the sense that, in coordinates close to the isolated structure, they take much larger values than at distances of approximately half the spacing between such structures. We also assume that we can choose a region $Y_m$ around the isolated structure such that all $f_{(s),m}$ have much smaller values on its boundaries and outside this region compared to the values inside, and such that all $Y_m$ do not overlap except at their boundaries. Under these assumptions, we may approximately consider that at each point in the coordinate space only one of the structures makes a significant contribution to the value of the function $H$:
\begin{eqnarray}\label{eq13_6}
H\left(\left\{\sum_m f_{(s),m}(X)\right\}_{s=1}^M \right) \approx H\left(f_{(1),m}(X), f_{(2),m}(X), ...,f_{(M),m}(X) \right) \Big|_{X \in Y_m}.
\end{eqnarray}
By repeating the derivations analogous to \eqref{eq13_2}--\eqref{eq13_5}, and assuming that the function $H\left(\left\{f_{(s),m}(X)\right\}_{s=1}^M\right)$ takes much larger values inside each region $Y_m$ than when approaching its boundaries, we obtain
\begin{eqnarray}\label{eq13_7}
\int_{\mathbb{R}^{dimX}} dX \ e^{iQ X} H\left(\left\{\sum_m f_{(s),m}(X)\right\}_{s=1}^M \right) \approx \nonumber \\
 \approx \left(\sum_m e^{iQX_m} \ \tilde{1}_{H(Y_m)} \right) \left(\int_{\mathbb{R}^{dimX}} dX \ e^{iQ \tilde{X}} H \left(\left\{f_{(s)}^{(1)}(\tilde{X})\right\}_{s=1}^M \right) \right),
\end{eqnarray}
where $\left\{f_{(s)}^{(1)}(\tilde{X})\right\}_{s=1}^M$ corresponds to a set of function $\left\{f_{(s),m}({X})\right\}_{s=1}^M$ for a single isolated structure located at the coordinate origin with respect to all variables in the set $X$.

\section{Calculation of potential-dependent functions}
In this section we will consider necessary for the differential scattering cross section functions which characterize the target. First of all, we need to know $\chi_0$-function for a single plane with uniformly distributed atoms. This function was calculated in \cite{PerPl1} and has the following form
\begin{eqnarray}\label{eq15}
\langle \chi_0^{(1)} \rangle= \pm A  \exp \left[-\frac{|x|}{R} \right],
\end{eqnarray}
where $A=2ZQ \alpha N_{p} \frac{R}{L_y}$, $\pm$ sign corresponds to opposite sing of incident particle charge, $Qe$ is the absolute value of the charge of incident particles. Let us consider the scattering problem for incident electrons with $Q=1$. 

The Fourier component of $\chi_0$-function for a single atom \cite{AIA96} is given by
\begin{eqnarray}\label{eq16}
\tilde{\chi}_{\vec{\mu}}= \frac{4 \pi Z \alpha}{\mu^2+R^{-2}},
\end{eqnarray}
where $\mu^2=\mu_x^2+\mu_y^2$.
Using \eqref{eq16} and the idea that $\langle  \chi_0^{(1)} \chi'^{ \ (1)}_0 \rangle$ for a plane is \linebreak $\langle  \chi_0^{(1)} \chi'^{ \ (1)}_0 \rangle =N_p \langle  \tilde{\chi}_0^{(1)} \tilde{\chi}'^{ \ (1)}_0 \rangle$, we can write $\langle  \chi_0^{(1)} \chi'^{ \ (1)}_0 \rangle$ through Fourier transformations 
\begin{eqnarray}\label{eq17}
\langle \chi_0^{(1)} \chi'^{ \ (1)}_0 \rangle = N_p \int_{-L_y/2}^{L_y/2}\frac{dy_0}{L_y} \int_{\mathbb{R}^4} \frac{d^2\mu}{(2 \pi)^2} \frac{d^2\mu'}{(2 \pi)^2} e^{i (\mu_x x+\mu'_x x')}e^{i (\mu_y (y-y_0)+\mu'_y (y'-y_0))} \tilde{\chi}_{\vec{\mu}} \tilde{\chi}_{\vec{\mu}'}.
\end{eqnarray}
Considering $L_y \gg R$, we can approximately write
\begin{eqnarray}\label{eq18}
\int_{-L_y/2}^{L_y/2}\frac{dy_0}{L_y} e^{-iy_0(\mu_y+\mu'_y)} \approx \int_{-\infty}^{\infty}\frac{dy_0}{L_y} e^{-iy_0(\mu_y+\mu'_y)},
\end{eqnarray}
which means
\begin{eqnarray}\label{eq19}
\int_{-L_y/2}^{L_y/2}\frac{dy_0}{L_y} e^{-iy_0(\mu_y+\mu'_y)} \approx  \frac{2 \pi}{L_y} \delta(\mu_y+\mu'_y).
\end{eqnarray}
So, after integration in right part of \eqref{eq17} over $y_0$ and $\mu'_y$ we receive 
\begin{eqnarray}\label{eq20}
\langle\chi_0^{(1)} \chi'^{ \ (1)}_0 \rangle = \frac{2 Z^2 \alpha^2 N_p}{\pi L_y}\int_{\mathbb{R}^3} d^2\mu d\mu'_x e^{i (\mu_x x+\mu'_x x')}e^{i \mu_y (y -y' )} \frac{1}{\mu_x^2+\mu_y^2+R^{-2}} \frac{1}{\mu_x^{' \ 2}+\mu_y^2+R^{-2}}.
\end{eqnarray}
In the previous section, we showed that all functions needed for calculation of \eqref{eq13} should be taken at $\Delta y=0$. If we  set $\Delta y=0$ and make all necessary integrations in \eqref{eq20} we obtain
\begin{eqnarray}\label{eq30}
\langle \chi_0^{(1)} \chi'^{ \ (1)}_0 \rangle \Big|_{\Delta y=0} = \pi^2 Z \alpha A \left\{ 1-\xi \left[L_{-1}(\xi)K_0(\xi)+L_0(\xi) K_1(\xi) \right] \right\},
\end{eqnarray}
where $\xi=\frac{|x|+|x'|}{R}$, $K_j (\xi)$ is modified Bessel function of the second kind, $L_j (\xi)$ is modified Struve function.

Also, we need to calculate $\langle (\chi_0^{(1)})^2  \rangle$, $\langle (\chi_0^{' \ (1)})^2  \rangle$. The formula for them is the same as \eqref{eq30} with the only difference in arguments. While $\langle \chi_0^{(1)} \chi'^{ \ (1)}_0 \rangle$ depends on $x$, $x'$, $\Delta y= y -y'$, $\langle (\chi_0^{(1)})^2 \rangle$ depends only on $x$, $y$. That means we need to make following substitutions in \eqref{eq17}: $x'=x$, $y'=y$ for $\langle (\chi_0^{(1)})^2  \rangle$. With $y'=y$, the difference $\Delta y$ becomes $\Delta y=0$. It mean that $\langle (\chi_0^{(1)})^2  \rangle$ coincides with $\langle \chi_0^{(1)} \chi'^{ \ (1)}_0 \rangle \Big|_{\Delta y=0} (x,x'=x)$:
\begin{eqnarray}\label{eq31}
\langle (\chi_0^{(1)})^2  \rangle = \pi^2 Z \alpha A \left\{ 1-\frac{2|x|}{R} \left[L_{-1}\left(\frac{2|x|}{R}\right) K_0\left(\frac{2|x|}{R}\right) + L_0\left(\frac{2|x|}{R}\right) K_1\left(\frac{2|x|}{R}\right) \right] \right\}.
\end{eqnarray}
The function $\langle(\chi_0^{' \ (1)})^2  \rangle$ is calculated in the same way but with substitutions $x=x'$, $y=y'$ in \eqref{eq17} which leads to $\langle (\chi_0^{' \ (1)})^2  \rangle=\langle (\chi_0^{(1)})^2  \rangle (x')$. Fig. \ref{fig_func} shows the form of potential-depentent functions $\langle \chi_0^{(1)} \rangle$, $\langle (\chi_0^{(1)})^2 \rangle$.

\begin{figure}[!h]
      \centering
	   \begin{subfigure}{0.49\linewidth}
		\includegraphics[width=\textwidth]{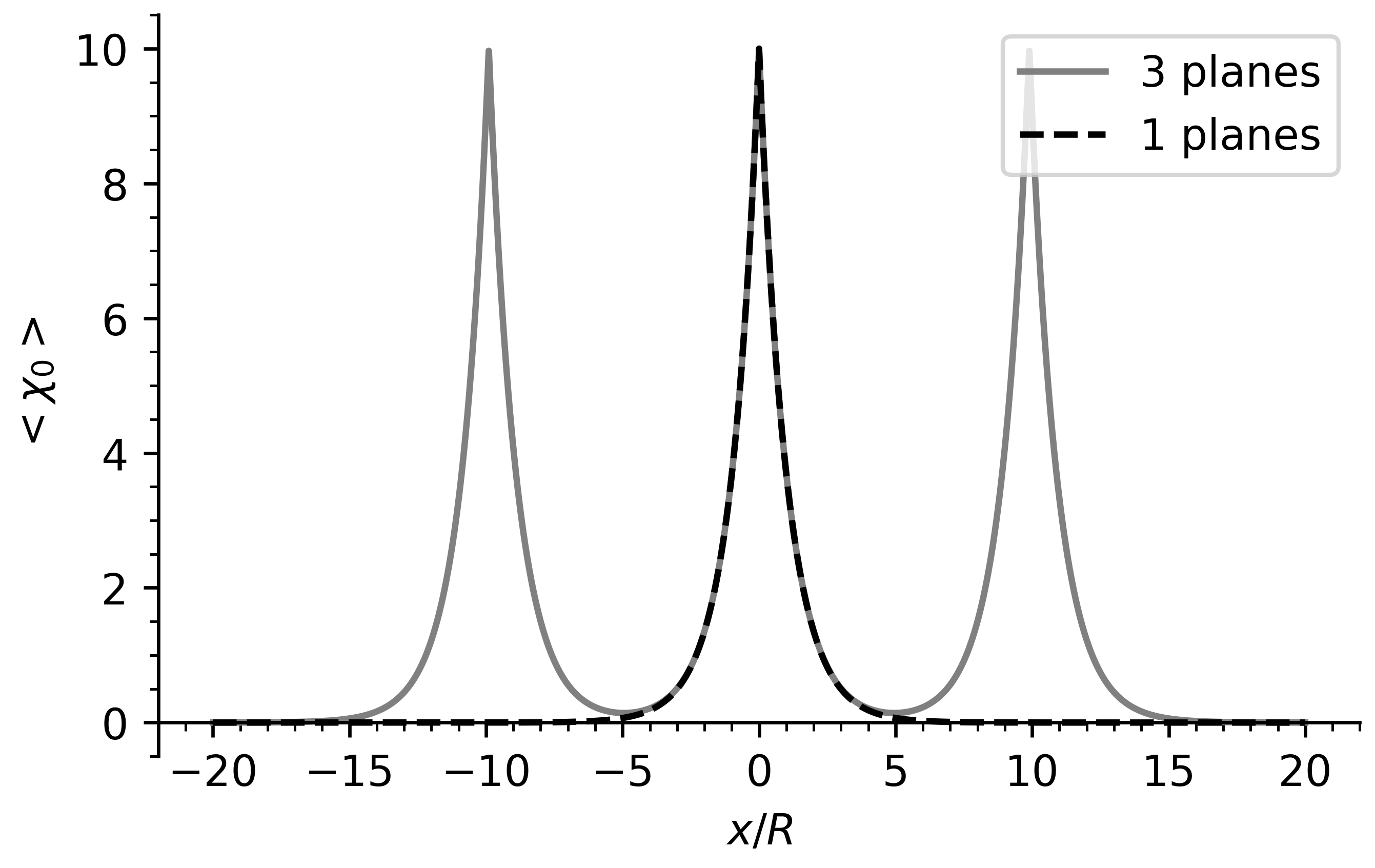}
		\caption{$\langle \chi_0^{(1)} \rangle (x)$}
		\label{fig:subfig_hi}
	   \end{subfigure}
	     \begin{subfigure}{0.49\linewidth}
		 \includegraphics[width=\textwidth]{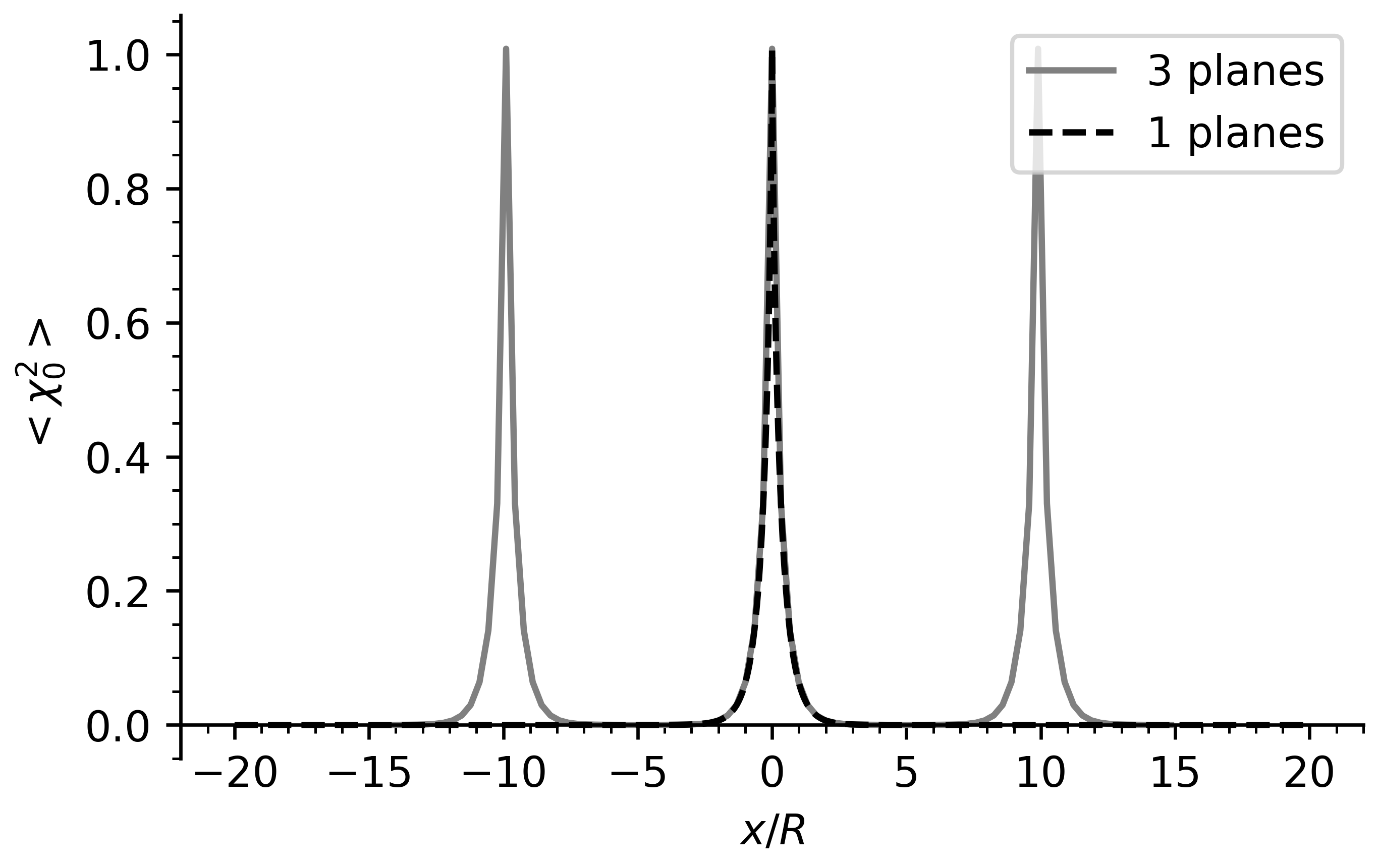}
		\caption{$ \langle (\chi_0^{(1)})^2 \rangle (x)$}
		 \label{fig:subfig_hi_sqv}
	      \end{subfigure}
\caption{Potential-dependent functions for $A=10$ for 1 and 3 atomic planes}
 \label{fig_func}
\end{figure}

\section{Calculation of the differential cross section of fast charged particles scattering on set of parallel atomic planes}

\begin{figure}[!h]
      \centering
	   \begin{subfigure}{0.48\linewidth}
		\includegraphics[width=\textwidth]{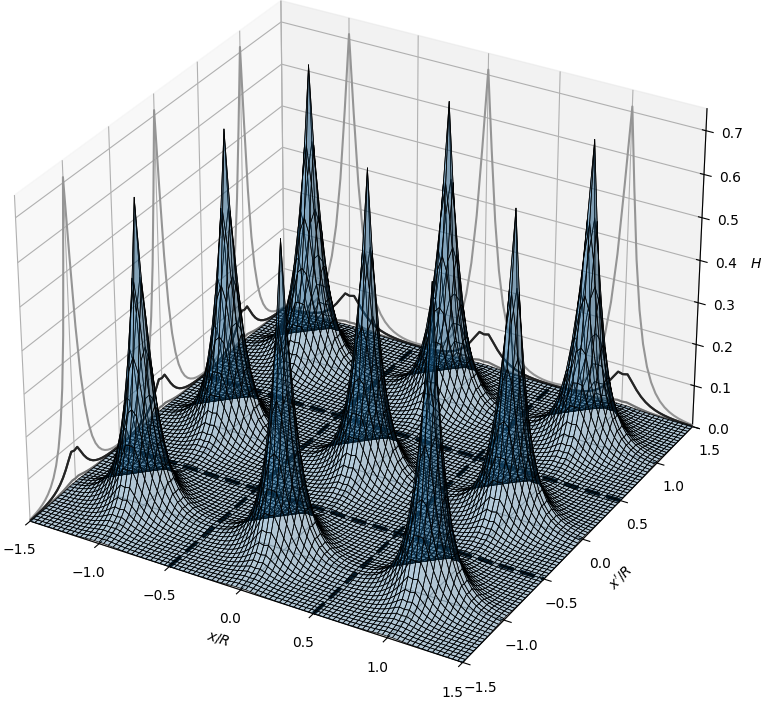}
		\caption{$H=\left(1-e^F \right)\left(1-e^{F'} \right)$}
		\label{fig:subfig_H0}
	   \end{subfigure}
	     \begin{subfigure}{0.48\linewidth}
		 \includegraphics[width=\textwidth]{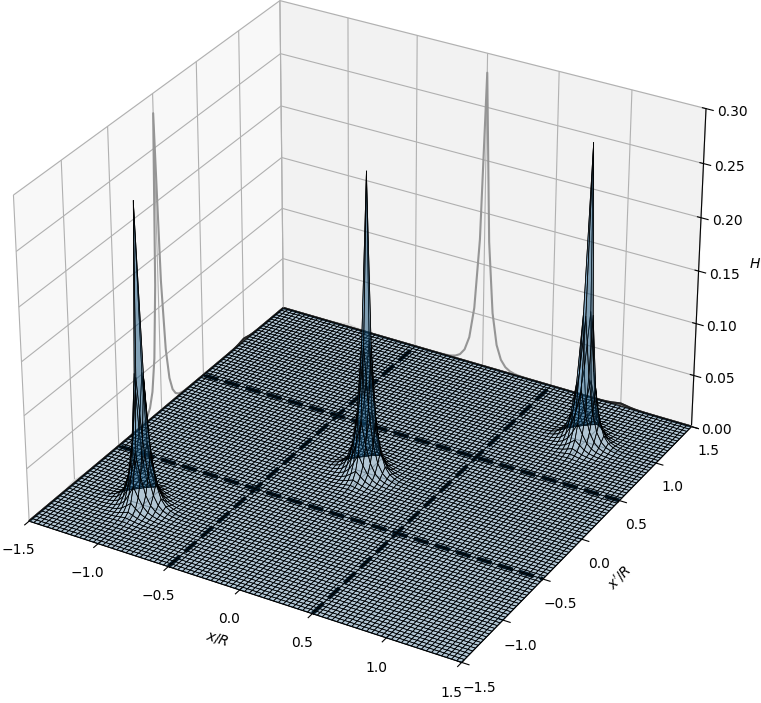}
		\caption{$H=\left(1-e^F \right)\left(1-e^{F'} \right)\left(1-e^{\tilde{G}} \right)$}
		 \label{fig:subfig_H1}
	      \end{subfigure}
\caption{Dependence of $H$ on $x$ and $x'$ for $A=1$ for 3 atomic planes}
 \label{fig_func}
\end{figure}

Let us consider the application of \eqref{eq13_7} to evaluate the integrals $\{I_j\}_{j=0}^3$, which define the differential cross section in \eqref{eq13}. Let the incident particles fall parallel to (100) planes of silicon. The distance between these planes is much larger than the screening radius of silicon atoms, so the atomic planes can be regarded as isolated. Physically, we have $N_x$ planes, but mathematically, formula \eqref{eq13_7} also works when adding "virtual" isolated structures. The variables $x$ and $x'$ refer to the same planes, but the integral $I_0$ includes all possible combinations of these planes, as illustrated in the Fig. \ref{fig:subfig_H0}. Mathematically, this is equivalent to performing the calculation for $N_x^2$ isolated structures located in the space of variables $(x,x')$.  

For the integrals $\{I_j\}_{j=0}^3$, we may take $X=\{x,x'\}$ and $Q=\{q_x,-q_x\}$. The index $m \in [1,N_x^2], \ m \in \mathbb{N}$, which enumerates the structures, can be conveniently replaced by indices $k,j \in [1,N_x], \ k,j \in \mathbb{N}$. We note that the enumeration scheme does not affect the result as long as it accounts for all structures contributing to the integral and counts each such structure exactly once. The functions $\langle \chi_0^{(1)} \rangle$, $\langle (\chi_0^{(1)})^2 \rangle$, $\langle \chi_0^{(1)} \chi'^{ \ (1)}_0 \rangle$ decay rapidly, so their finite sums -- the functions $F$, $F'$, $\tilde{G}$ -- exhibit the same properties as $\sum_m f_{(s),m}$: one can approximately consider that at each point in the coordinate space $(x,x')$ only one structure gives the main contribution to these functions. Moreover, the contribution of $m$-th function becomes sufficiently small at a distance of approximately $a/2$ from $m$-th structure and can be neglected. Therefore, the regions $Y_m$ can be limited by the lines $x = x_m \pm a/2$, $x' = x_m \pm a/2$ in the same way as shown in the Fig. \ref{fig_func} with dashed lines. The functions $F$, $F'$, $\tilde{G}$ are then defined as
\begin{eqnarray}\label{eq32}
F=\sum_{j=1}^{N_x} F_j, \ F_j(x)=i \langle \chi_{(j)}\rangle -\frac{1}{2}  \langle  \chi_{(j)}^2 \rangle, 
\end{eqnarray}
\begin{eqnarray}\label{eq33}
F'=\sum_{j=1}^{N_x} F'_j, \ F'_j(x')=F_j^*(x'), 
\end{eqnarray}
\begin{eqnarray}\label{eq34}
\tilde{G}=\sum_{j,k=1}^{N_x} \tilde{G}_{jk} \delta_{jk}, \ \tilde{G}_{jk}=\langle  \chi_{(j)} \chi'_{(k)} \rangle \Big|_{\Delta_y=0}, 
\end{eqnarray}
where $\delta_{jk}$ is the Kronecker symbol, which appears in the definition of  $\tilde{G}$ because in the differential cross section it appears as $\tilde{G}(x,x')=\sum_{j=1}^{N_x} \langle  \chi_{(j)} \chi'_{(j)} \rangle \Big|_{\Delta_y=0}$.

For the integral $I_0$, the function $H$ is defined as $H\left(\sum_{j=1}^{N_x} F_j, \sum_{k=1}^{N_x} F'_k \right)=(1-e^F)(1-e^{F'})$. This expression is constructed in the way that when $F$ або $F'$ are equal to $0$, the exponential of the corresponding function becomes equal to 1, and the entire bracket of the form $(1-e^F)=0$. It is sufficient for one of brackets to be $0$ to lead to $H=0$. If in each coordinate point, $H\left(\sum_{j=1}^{N_x} F_j (x), \sum_{k=1}^{N_x} F'_k (x') \right) \approx H\left( F_j (x), F'_k (x') \right) \Big|_{\{x,x'\} \in Y_{jk}}$,
then according to \eqref{eq13_7}
\begin{eqnarray}\label{eq35}
I_0=D_{N_x} I_0^{(1)}, 
\end{eqnarray}
where $D_{N_x}=\Big| \sum_{k=1}^{N_x} e^{i {q}_{x}x_k} \Big|^2$, $I_0^{(1)}=\int_{\mathbb{R}^2}dx dx' \ e^{i{q}_{x}(x-x')} \left(1-e^{F^{(1)}}\right) \left(1-e^{F^{' \ (1)}} \right)$, \linebreak $F^{(1)}(x)=i \langle \chi^{(1)}_0\rangle -\frac{1}{2} \langle  (\chi^{(1)}_0)^2 \rangle$, $F^{' \ (1)}(x')=(F^{(1)})^*(x')$.
We note that $D_{N_x}$ is invariant with respect to shift of the coordinate system, so, we do not need to specify nothing more than distance $a$ and the number of planes $N_x$.

In the remaining integrals $\{I_j\}_{j=1}^3$, unlike $I_0$, there is a factor $(1-e^{\tilde{G}})$. The function $H$ for these integrals is defined as follows: for $I_1$, $H = \left(1-e^F \right)\left(1-e^{F'} \right)\left(1-e^{\tilde{G}} \right)$; for $I_2$, \linebreak $H = \left(1-e^{\tilde{G}} \right)$; and for $I_3$, $H = \left(1-e^F \right)\left(1-e^{\tilde{G}} \right)$. Unlike the brackets $\left(1-e^F \right)\left(1-e^{F'} \right)$, which are products of functions $\left(1-e^F \right)$ and $\left(1-e^{F'} \right)$ of the corresponding independent variables, the function $\tilde{G}$ ``ties together’’ the variables $x$ and $x'$ because, by definition, only the diagonal elements $\tilde{G}_{jj}$ are nonzero. This means that a nonzero contribution arises only from regions where $x$ and $x'$ are within the influence of the same structure. If $x$ and $x'$ are within the regions of different structures, the corresponding integrals for such $x$ and $x'$ are zero. A characteristic shape of such $H$ is illustrated in the Fig. \ref{fig:subfig_H1} using as example $H = \left(1-e^F \right)\left(1-e^{F'} \right)\left(1-e^{\tilde{G}} \right)$. In this situation, $\tilde{1}_{H(Y_{jk})} = \delta_{jk}$ and $\sum_{j,k=1}^{N_x} e^{iq_x(x_j-x_k)} \ \tilde{1}_{H(Y_{jk})} = N_x$. Then
\begin{eqnarray}\label{eq36}
I_j={N_x} I_j^{(1)}, \ j=1,2,3; 
\end{eqnarray}
where integrals $\{I^{(1)}_j \}_{j=1}^3$ are $I_1^{(1)}=\int_{\mathbb{R}^2}dx dx' \ e^{i{q}_{x}(x-x')} \left(1-e^{F^{(1)}} \right) \left(1-e^{F^{' \ (1)}} \right)\left(1-e^{\tilde{G}^{(1)}} \right)$, \linebreak $I_2^{(1)}=\int_{\mathbb{R}^2}dx dx' \ e^{i{q}_{x}(x-x')} \left(1-e^{\tilde{G}^{(1)}} \right)$, $I_3^{(1)}=2Re\int_{\mathbb{R}^2}dx dx' \ e^{i{q}_{x}(x-x')} \left(1-e^{F^{(1)}} \right)\left(1-e^{\tilde{G}^{(1)}} \right)$, $\tilde{G}^{(1)}(x,x')=\langle  \chi^{(1)}_0 \chi'^{ \ (1)}_0 \rangle \Big|_{\Delta_y=0}$.
Considering \eqref{eq35}, \eqref{eq36}, formula \eqref{eq13} becomes
\begin{eqnarray}\label{eq37}
\frac{d\sigma}{dq_{x}}=\frac{L_y}{2\pi} \left\{ D_{N_x} I_0^{(1)}+N_x \left( -I_1^{(1)} - I_2^{(1)} +I_3^{(1)} \right)  \right\}.
\end{eqnarray}
To obtain the final result we performed numerical calculations of formulas \eqref{eq12_1} and \eqref{eq37}. In formula \eqref{eq12_1} we made less approximations but formula \eqref{eq37} is easier and faster to compute. 

\section{Discussion}
\begin{figure}[!h]
      \centering
	   \begin{subfigure}{0.49\linewidth}
		\includegraphics[width=\textwidth]{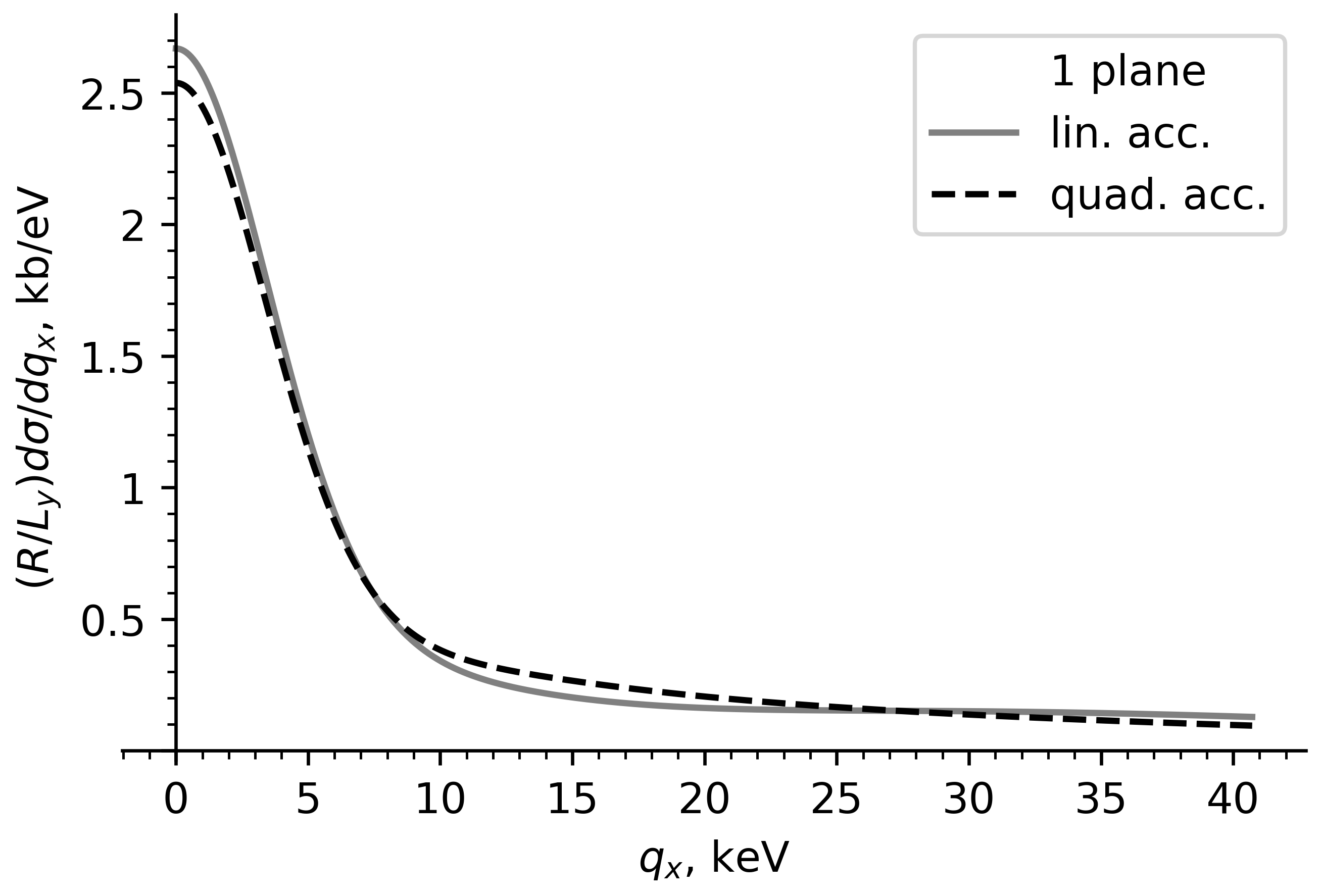}
		\caption{accuracy up to linear and quadratic potential terms}
		\label{fig:subfig_1pl_lq0}
	   \end{subfigure}
\vfill
\begin{subfigure}{0.48\linewidth}
		 \includegraphics[width=\textwidth]{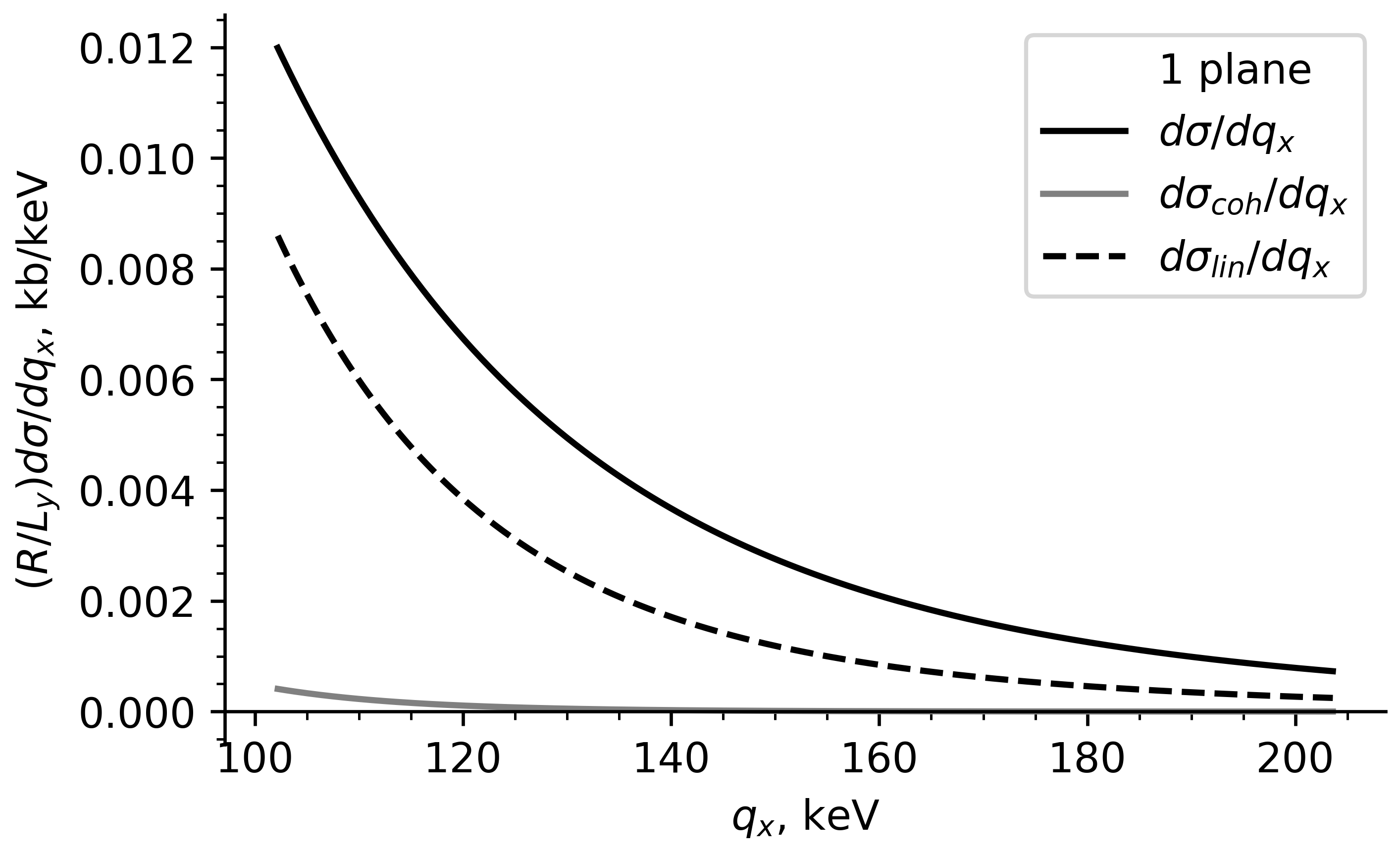}
		\caption{$\frac{d\sigma}{dq_x}$, $\frac{d\sigma_{coh}}{dq_x}$, $\frac{d\sigma_{lin}}{dq_x}$}
		 \label{fig:subfig_1pl_cil}
	      \end{subfigure}
	     \begin{subfigure}{0.48\linewidth}
		 \includegraphics[width=\textwidth]{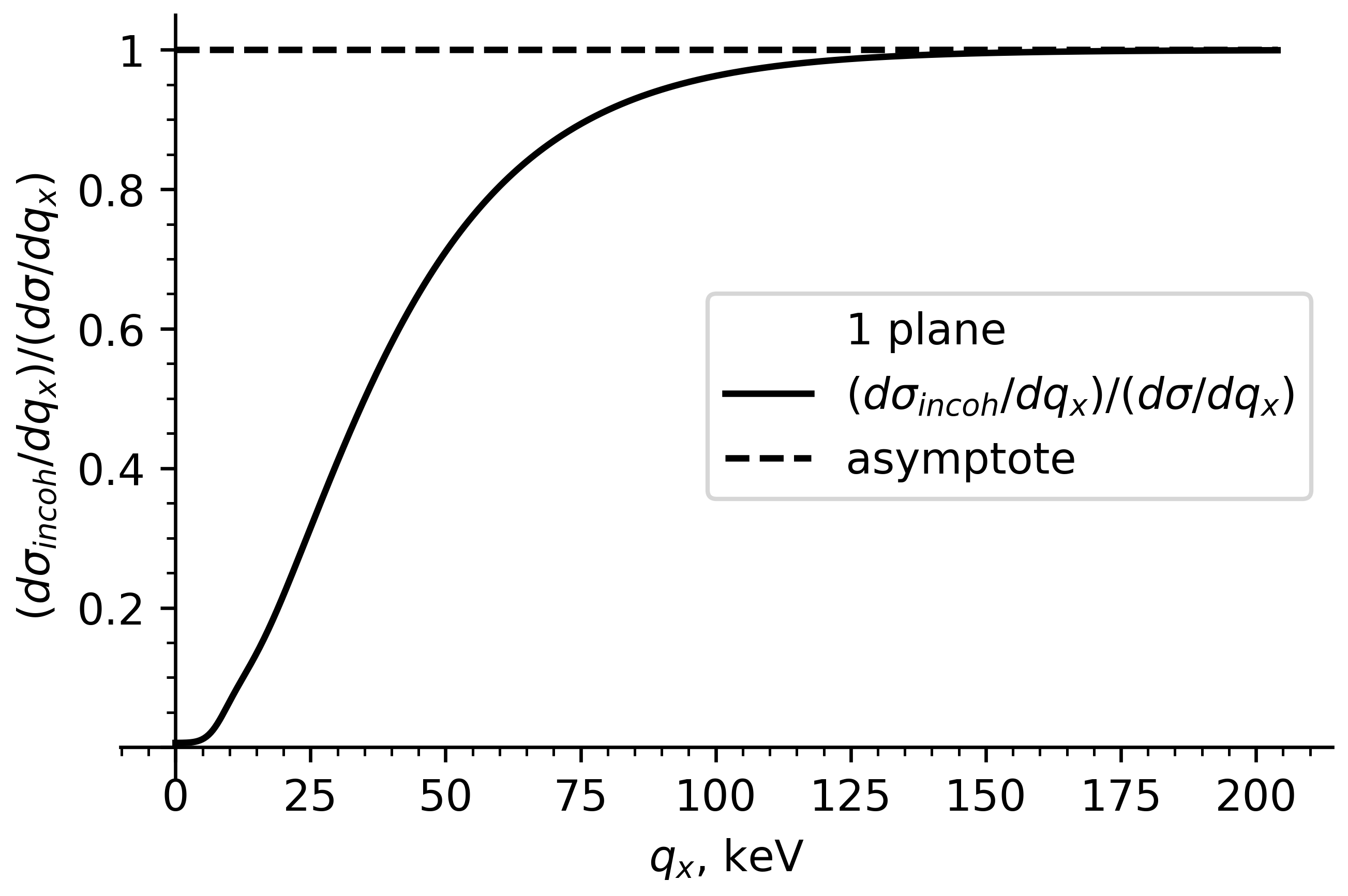}
		\caption{$\frac{d\sigma_{incoh}}{dq_x}/\frac{d\sigma}{dq_x}$}
		 \label{fig:subfig_1pl_ci}
	      \end{subfigure}
\caption{The differential cross section of fast charged particles scattering on 1 atomic plane. $\frac{d\sigma_{lin}}{dq_x}$ denotes the differential scattering cross section calculated with accuracy up to linear potential terms}
 \label{fig_pl1}
\end{figure}

\begin{figure}[!h]
      \centering
	   \begin{subfigure}{0.49\linewidth}
		\includegraphics[width=\textwidth]{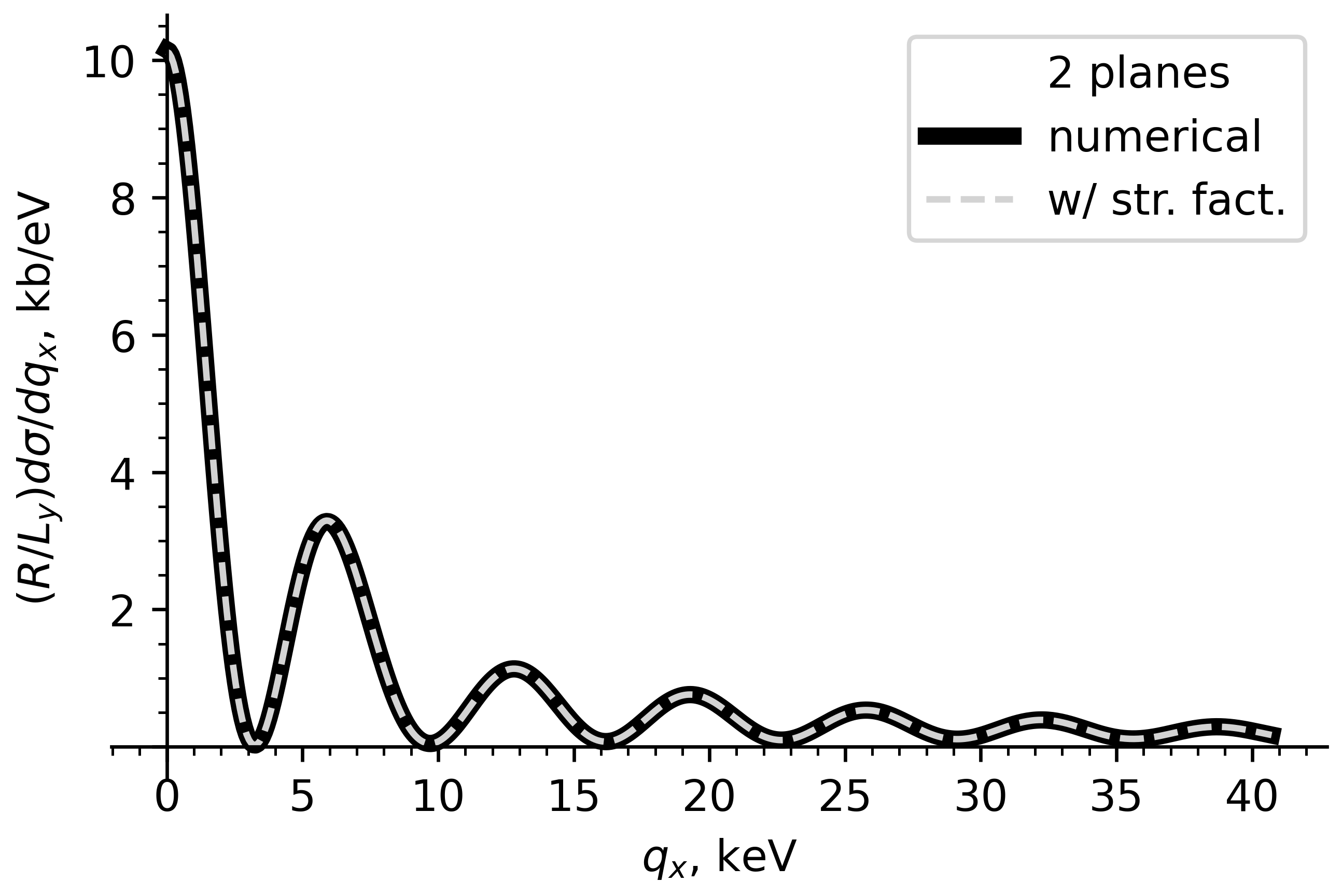}
		\caption{numerical integration and structure factors approaches}
		\label{fig:subfig_2pl_ns0}
	   \end{subfigure}
	     \begin{subfigure}{0.49\linewidth}
		 \includegraphics[width=\textwidth]{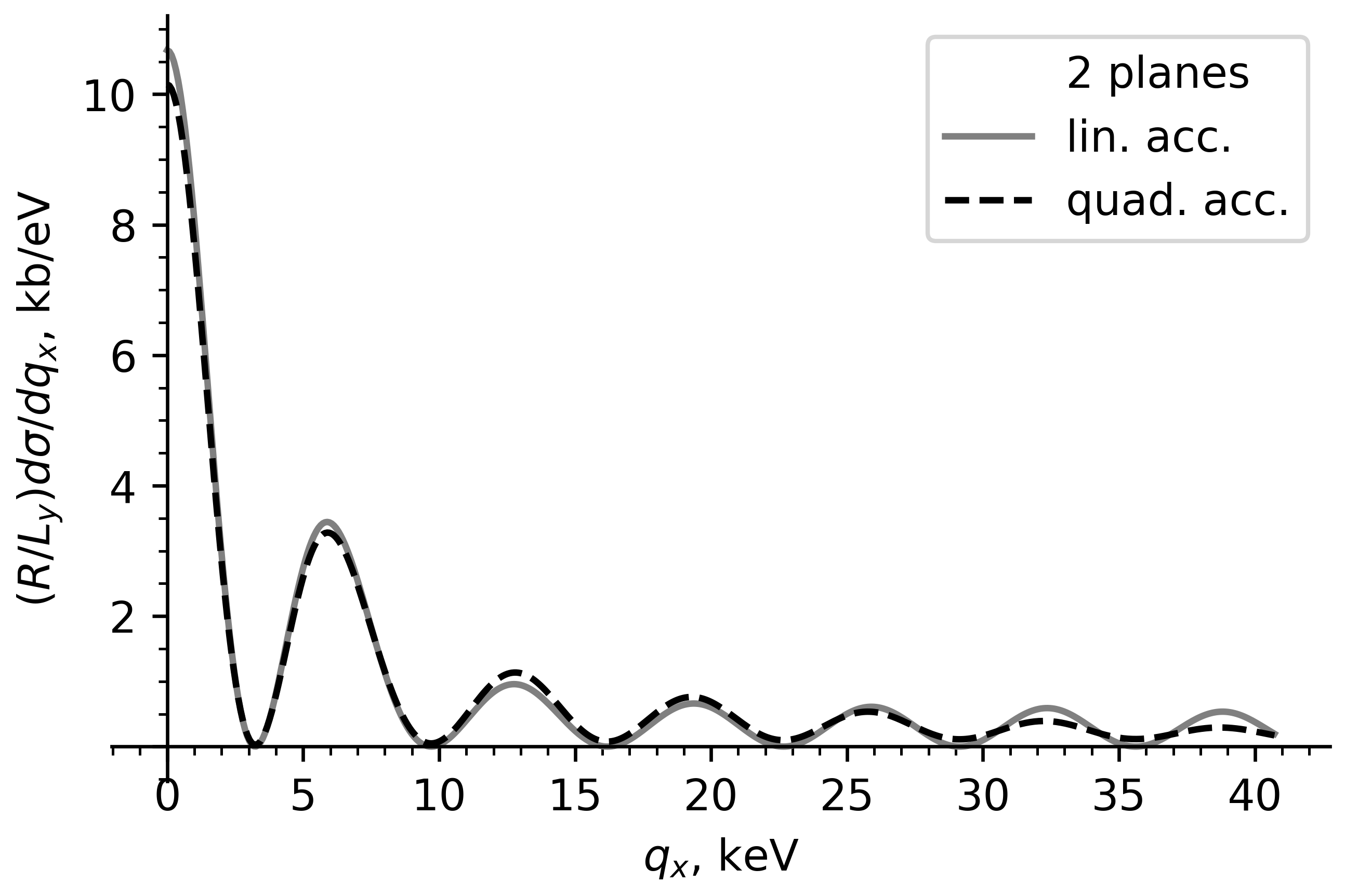}
		\caption{accuracy up to linear and quadratic potential terms}
		 \label{fig:subfig_2pl_lq0}
	      \end{subfigure}
\caption{The differential cross section of fast charged particles scattering on 2 atomic planes}
 \label{fig_pl2}
\end{figure}

\begin{figure}[ht]
      \centering
\begin{subfigure}{0.49\linewidth}
\centering
		\includegraphics[width=\textwidth]{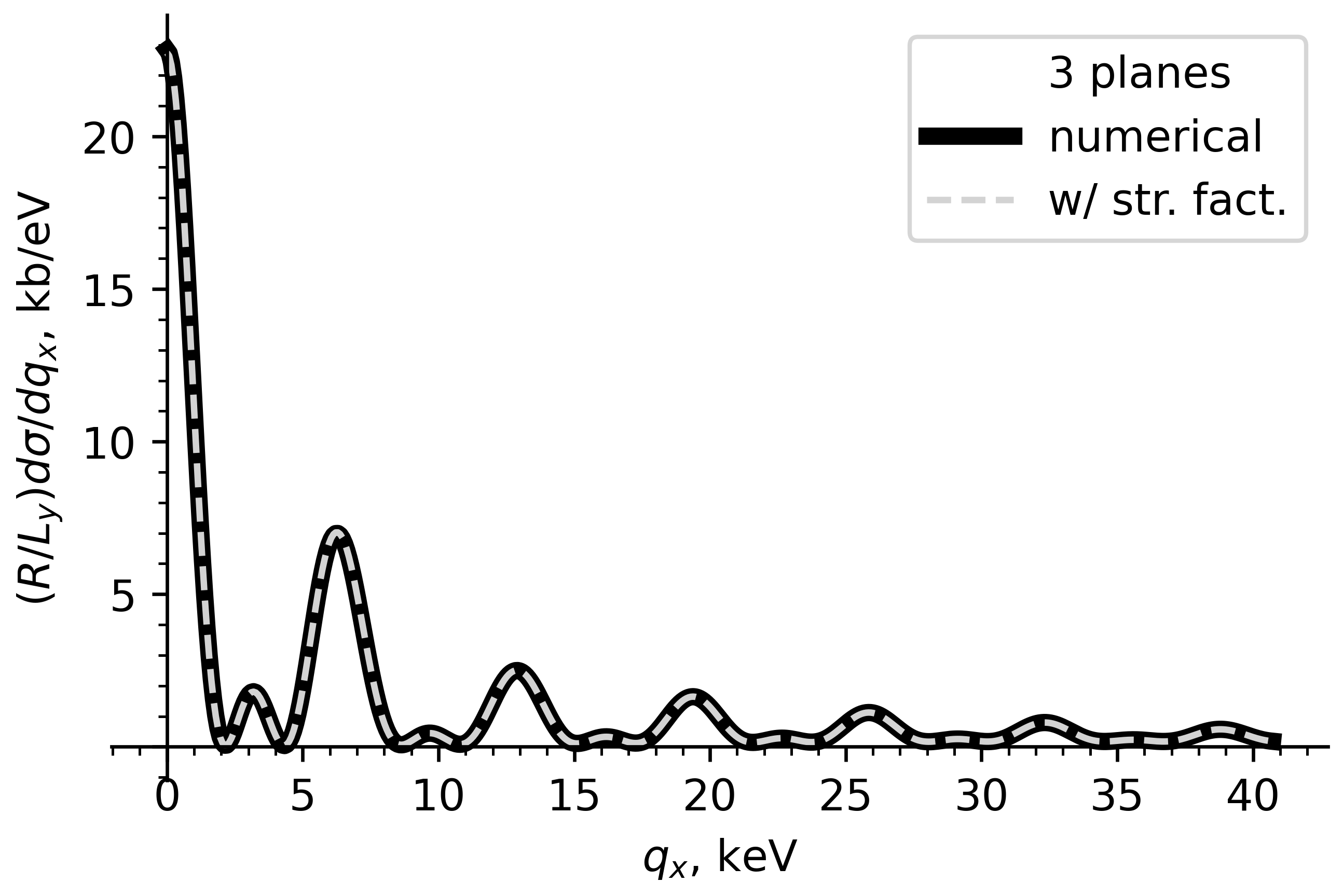}
		\caption{numerical integration and structure factors approaches, $q_x \in [0, 40.8]$ keV}
		\label{fig:subfig_3pl_ns0}
	    \end{subfigure}
	   \begin{subfigure}{0.49\linewidth}
		\includegraphics[width=\textwidth]{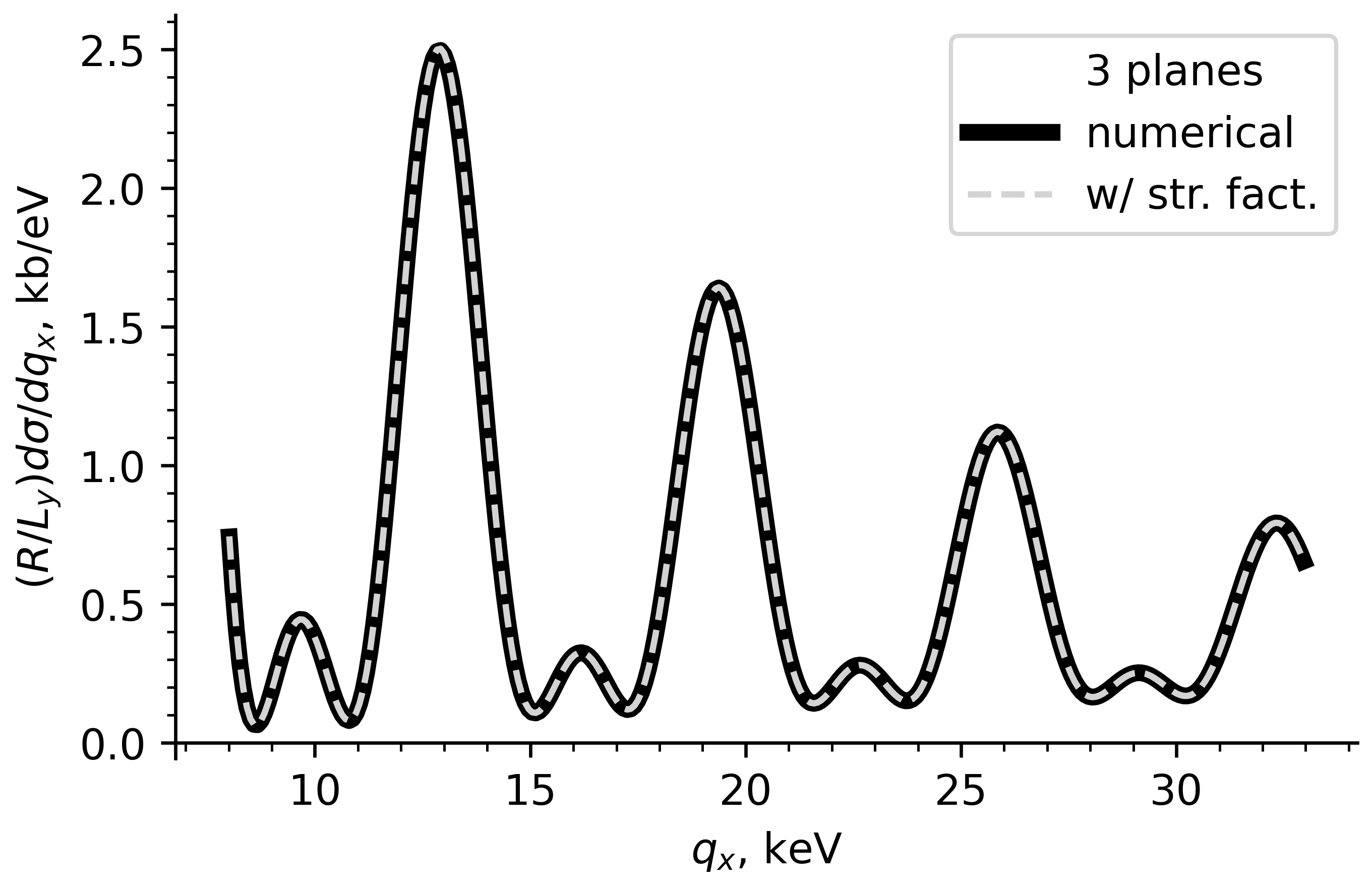}
		\caption{numerical integration and structure factors approaches, $q_x \in [8, 33]$ keV}
		\label{fig:subfig_2pl_ns2}
	   \end{subfigure}
\vfill
	     \begin{subfigure}{0.48\linewidth}
		 \includegraphics[width=\textwidth]{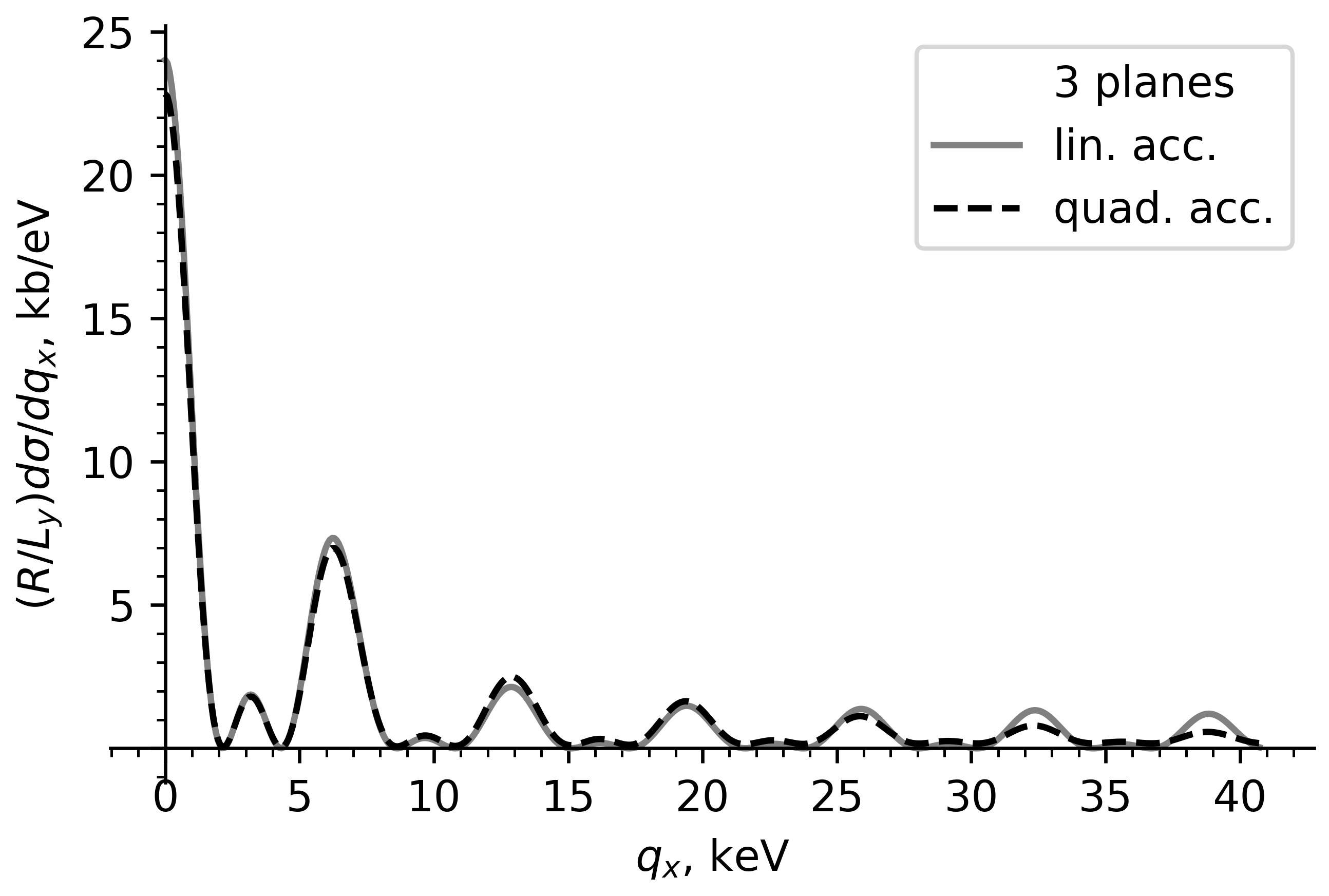}
		 \caption{accuracy up to linear and quadratic potential terms, $q_x \in [0, 40.8]$ keV}
		 \label{fig:subfig_2pl_lq0}
	      \end{subfigure}
\begin{subfigure}{0.48\linewidth}
		 \includegraphics[width=\textwidth]{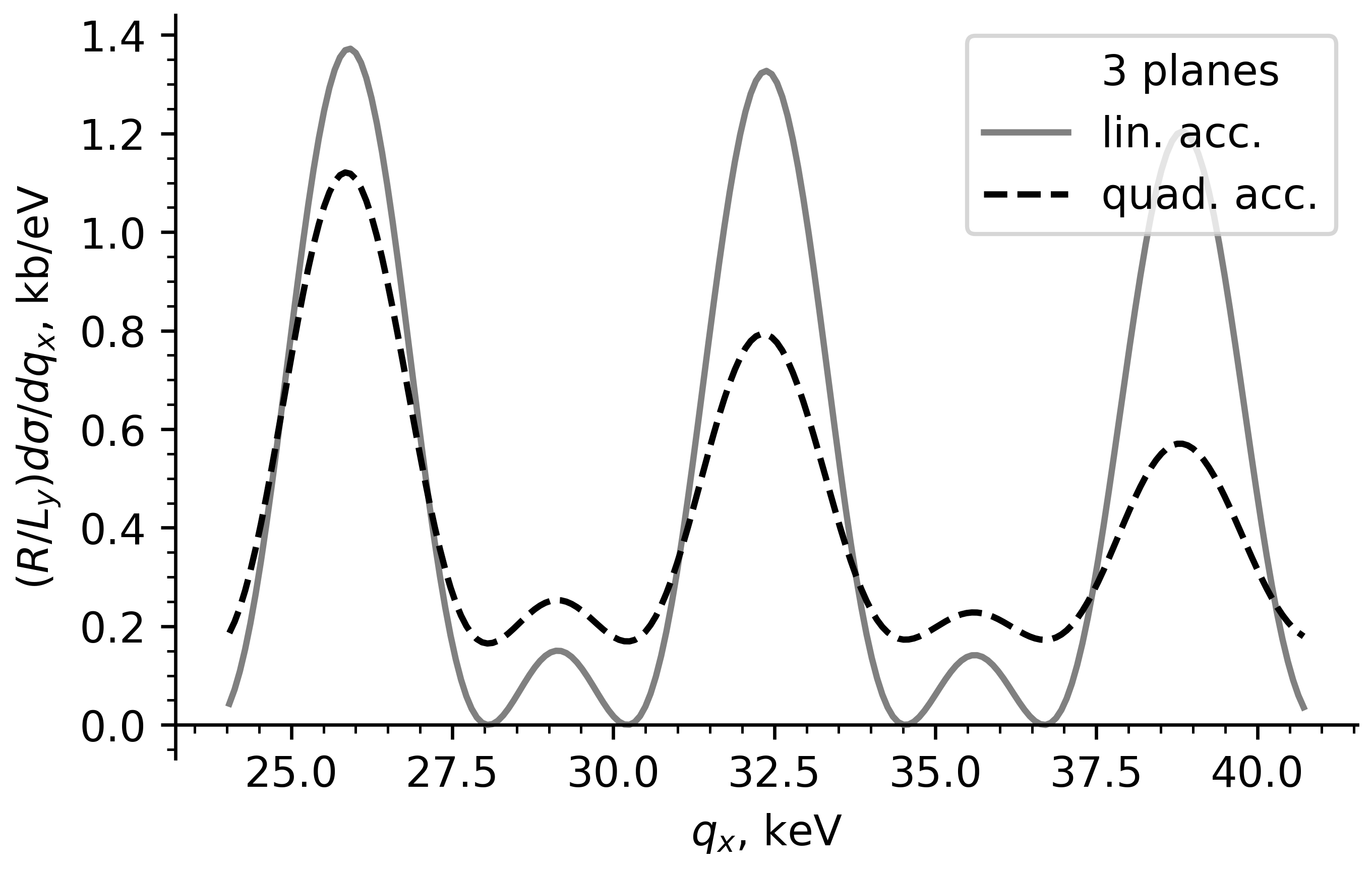}
		 \caption{accuracy up to linear and quadratic potential terms, $q_x \in [24, 40.8]$ keV}
		 \label{fig:subfig_2pl_lq3}
	      \end{subfigure}
\caption{The differential cross section of fast charged particles scattering on 3 atomic planes}
 \label{fig_pl3}
\end{figure}

\begin{figure}[ht]
      \centering
\begin{subfigure}{0.49\linewidth}
\centering
		\includegraphics[width=\textwidth]{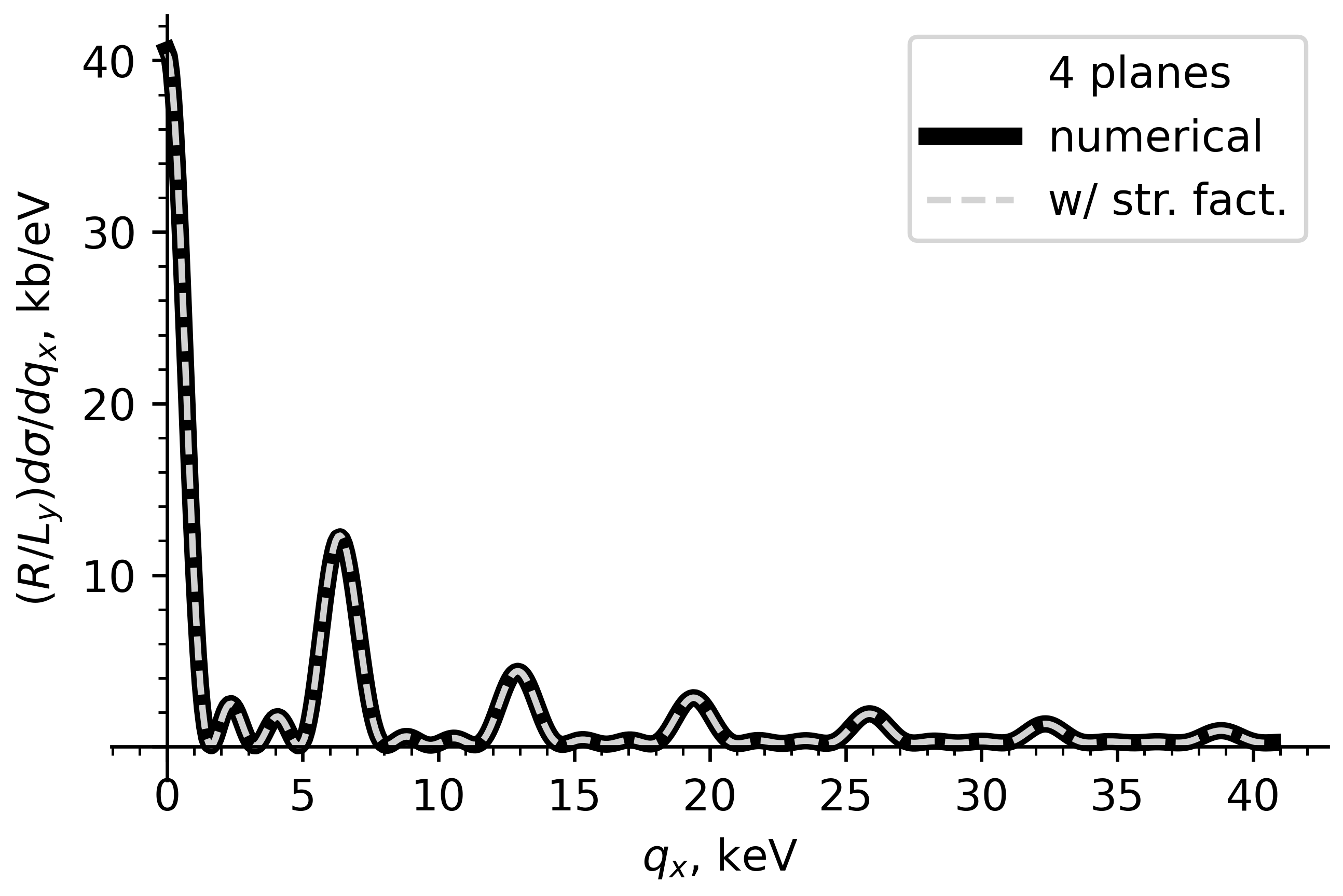}
		\caption{numerical integration and structure factors approaches, $q_x \in [0, 40.8]$ keV}
		\label{fig:subfig_4pl_ns0}
	    \end{subfigure}
	   \begin{subfigure}{0.49\linewidth}
		\includegraphics[width=\textwidth]{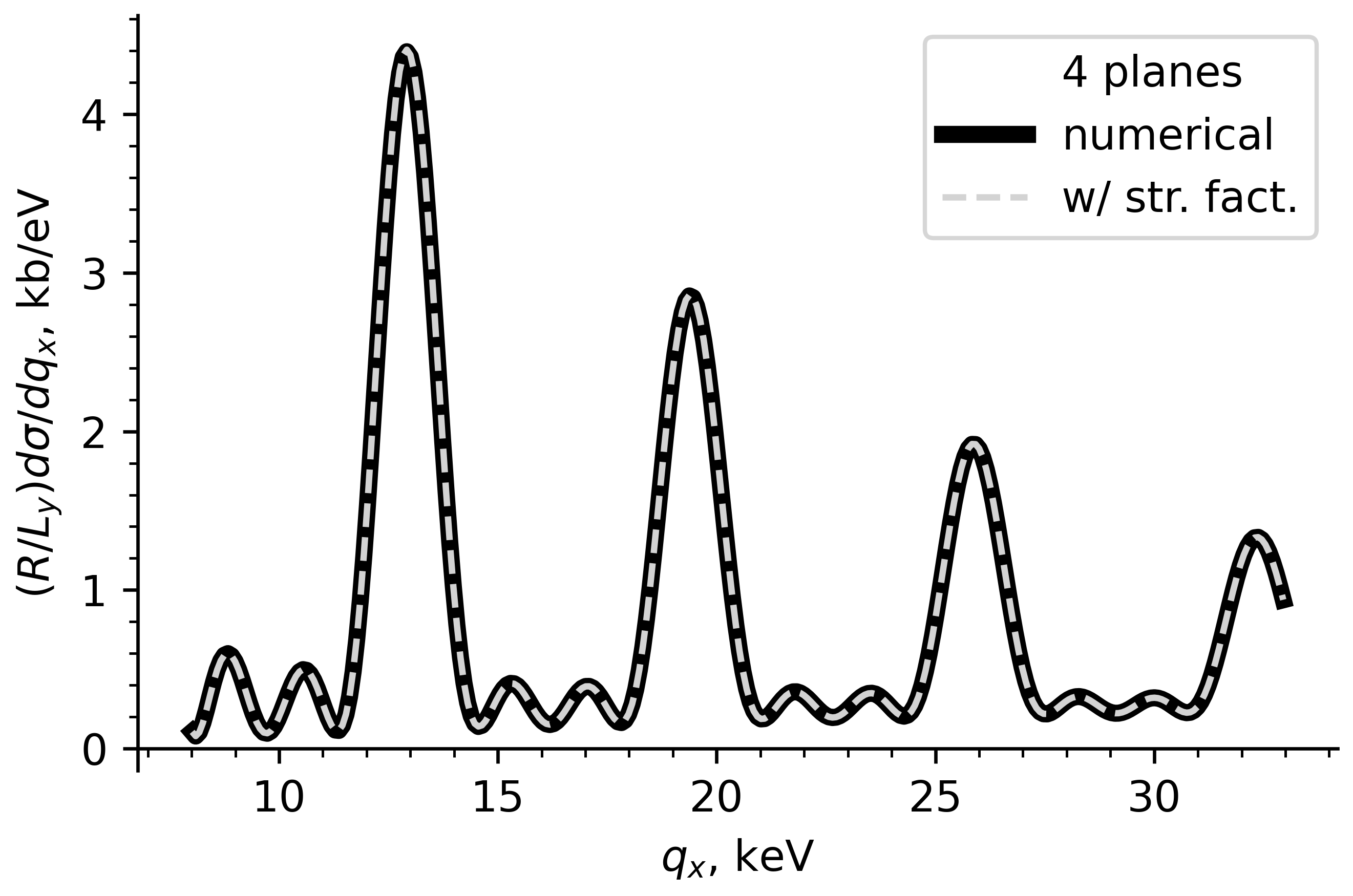}
		\caption{numerical integration and structure factors approaches, $q_x \in [8, 33]$ keV}
		\label{fig:subfig_4pl_ns2}
	   \end{subfigure}
\vfill
\begin{subfigure}{0.49\linewidth}
		 \includegraphics[width=\textwidth]{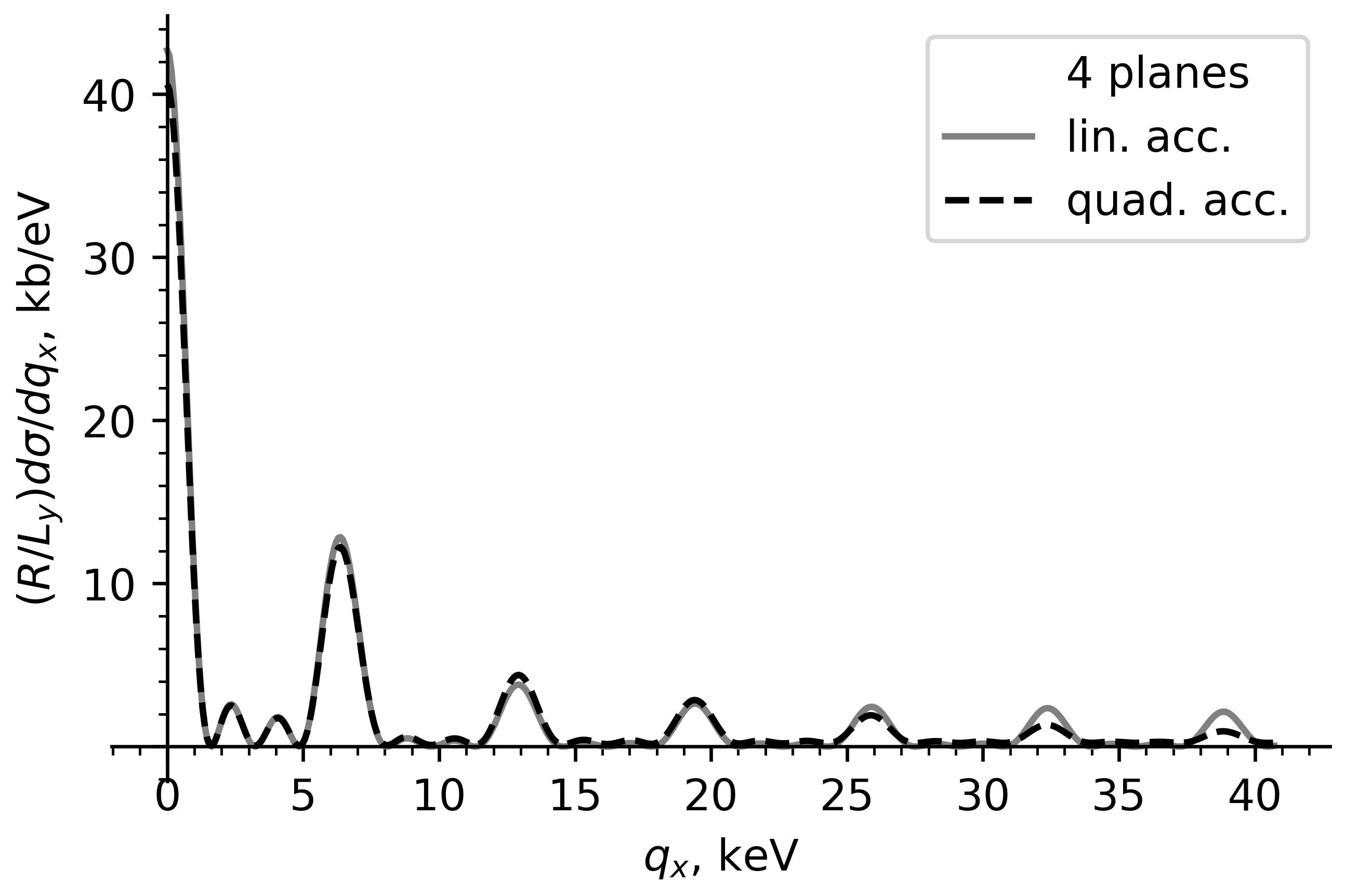}
		 \caption{accuracy up to linear and quadratic potential terms, $q_x \in [0, 40.8]$ keV}
		 \label{fig:subfig_4pl_lq0}
	      \end{subfigure}
	     \begin{subfigure}{0.49\linewidth}
		 \includegraphics[width=\textwidth]{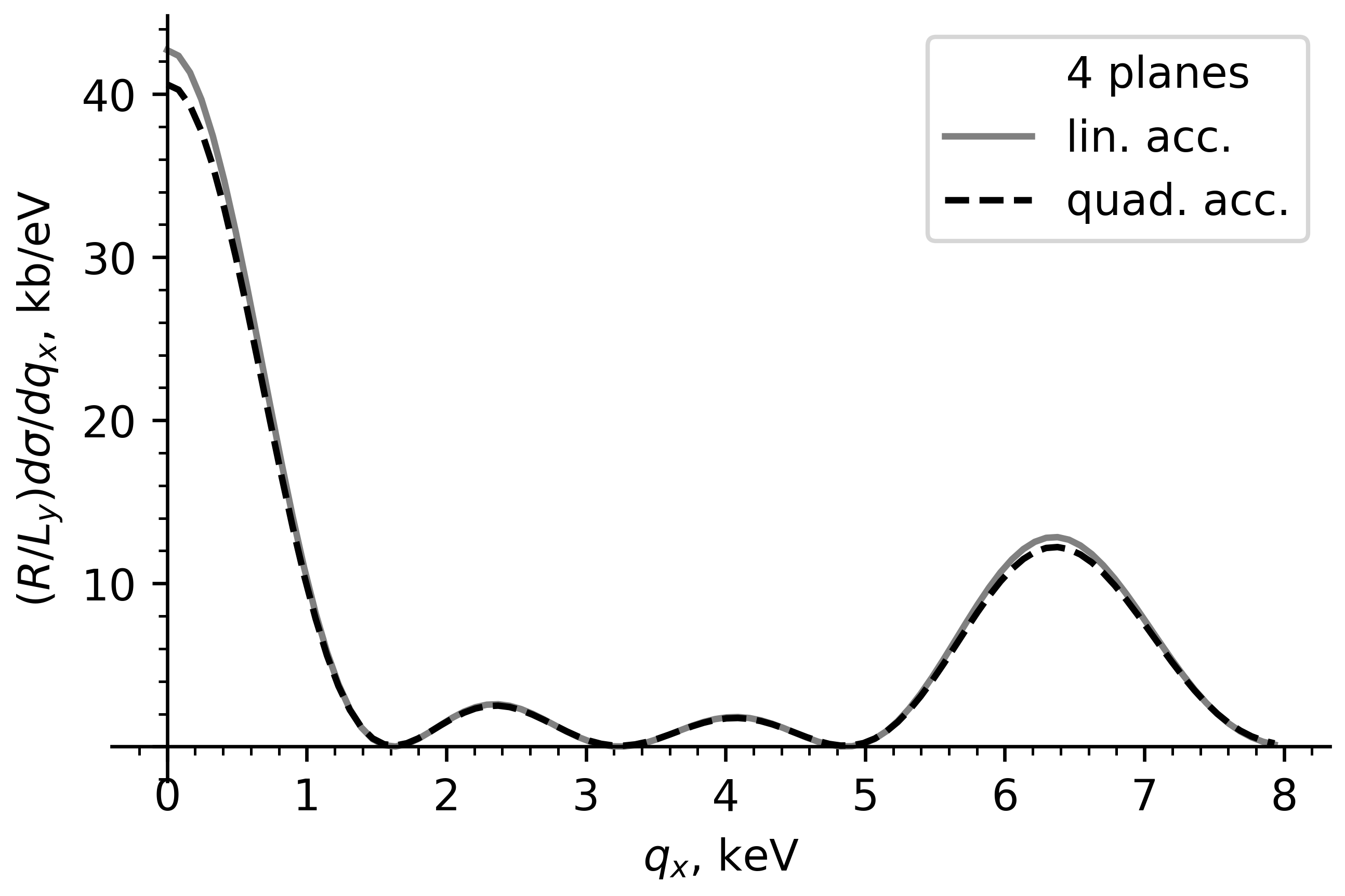}
		 \caption{accuracy up to linear and quadratic potential terms, $q_x \in [0, 8]$ keV}
		 \label{fig:subfig_4pl_lq1}
	      \end{subfigure}

\vfill

 \begin{subfigure}{0.49\linewidth}
		 \includegraphics[width=\textwidth]{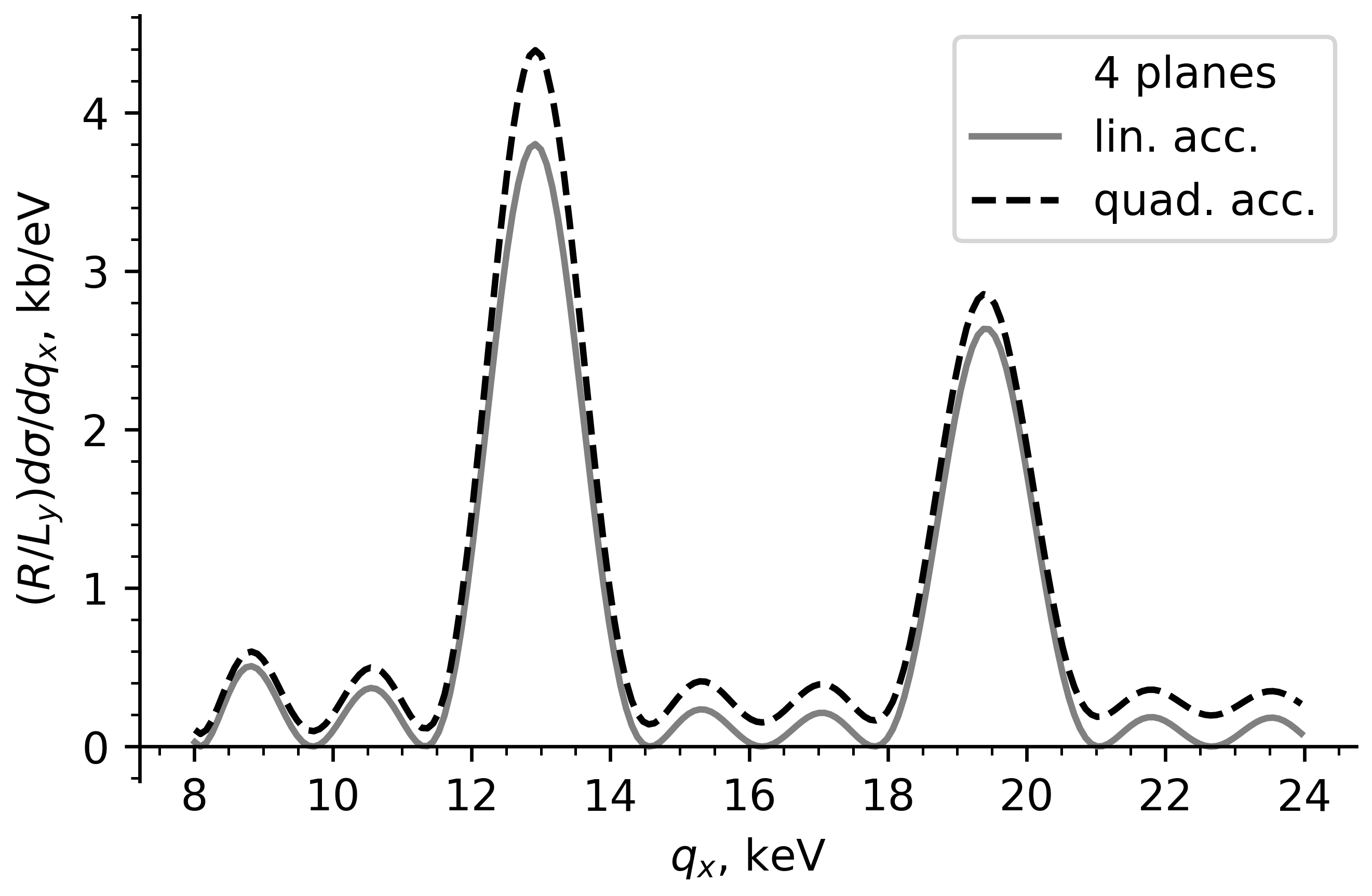}
		 \caption{accuracy up to linear and quadratic potential terms, $q_x \in [8, 24]$ keV}
		 \label{fig:subfig_4pl_lq2}
	      \end{subfigure}
\begin{subfigure}{0.49\linewidth}
		 \includegraphics[width=\textwidth]{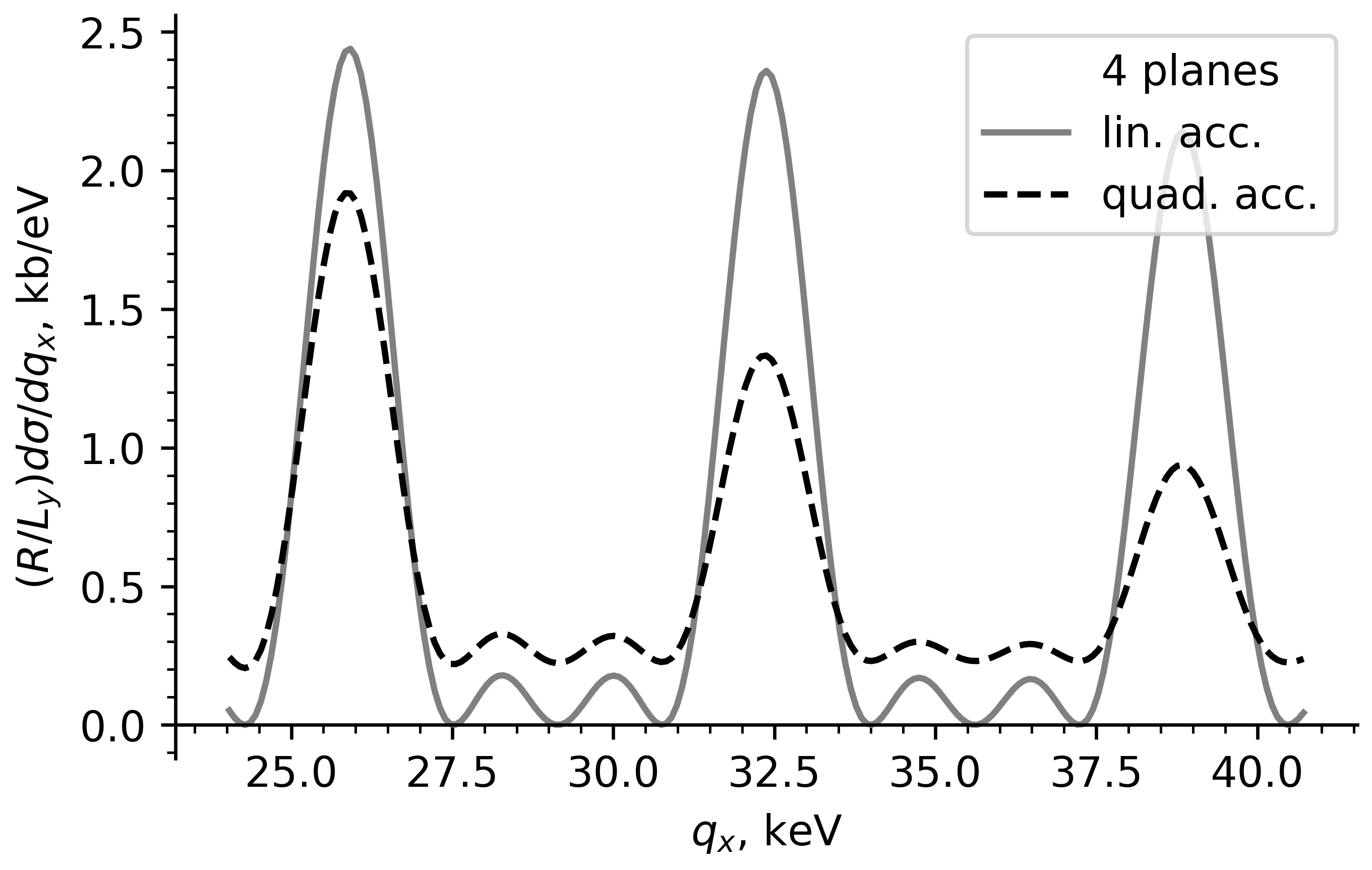}
		 \caption{accuracy up to linear and quadratic potential terms, $q_x \in [24, 40.8]$ keV}
		 \label{fig:subfig_4pl_lq3}
	      \end{subfigure}
\caption{The differential cross section of fast charged particles scattering on 4 atomic planes}
 \label{fig_pl4}
\end{figure}

\begin{figure}[ht]
      \centering
\begin{subfigure}{0.49\linewidth}
\centering
		\includegraphics[width=\textwidth]{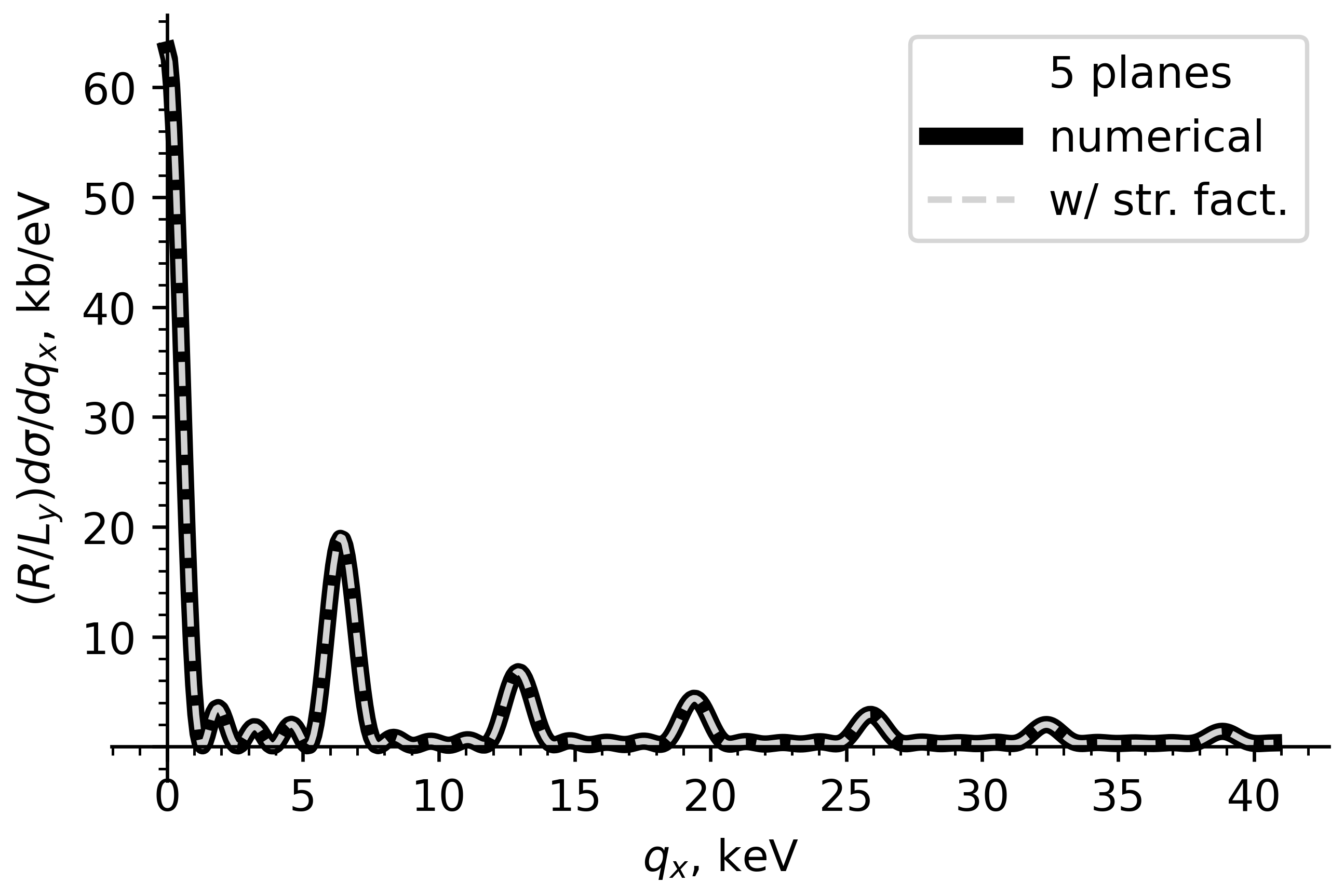}
		\caption{numerical integration and structure factors approaches, $q_x \in [0, 40.8]$ keV}
		\label{fig:subfig_5pl_ns0}
	    \end{subfigure}
	   \begin{subfigure}{0.49\linewidth}
		\includegraphics[width=\textwidth]{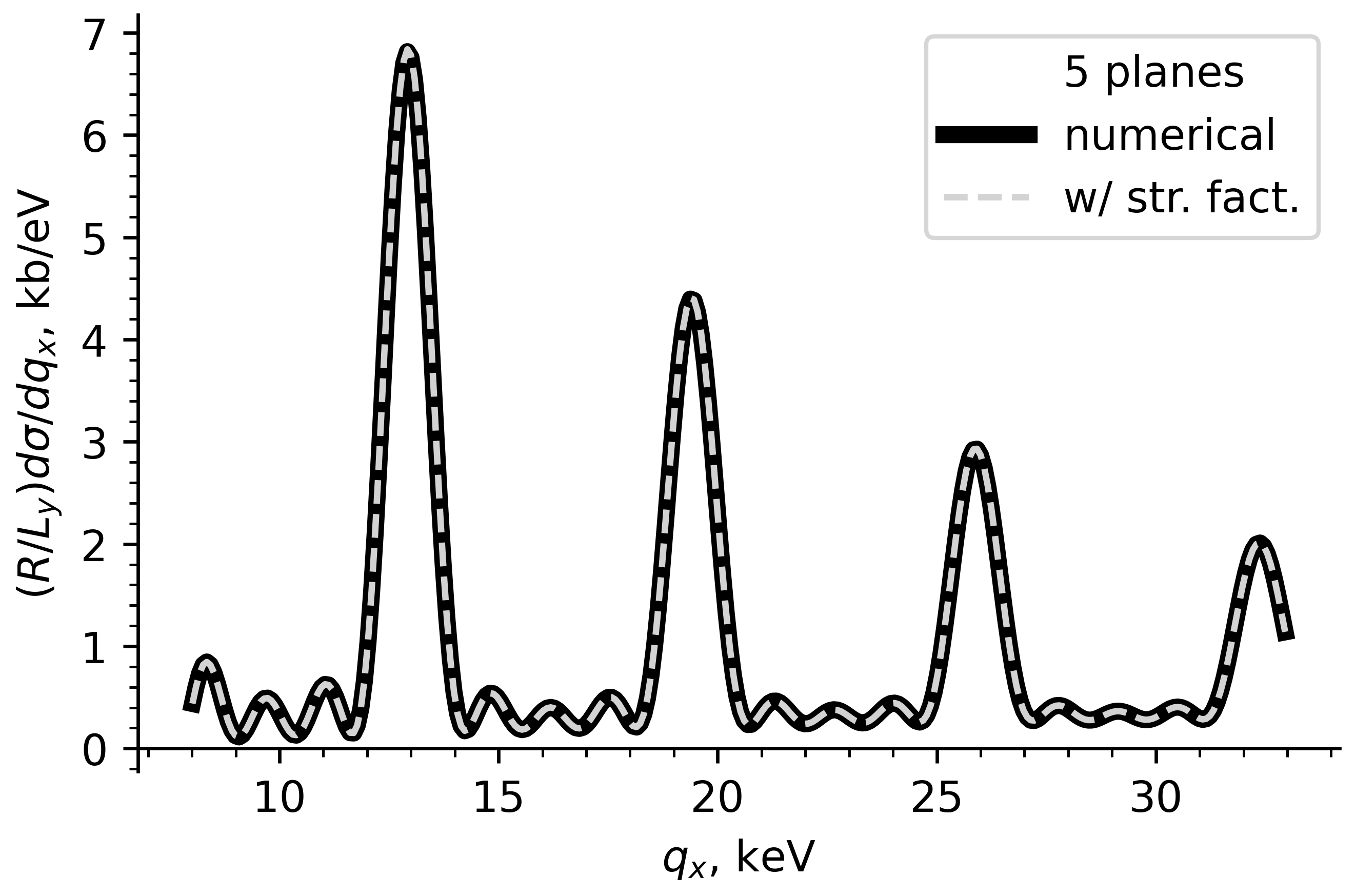}
		\caption{numerical integration and structure factors approaches, $q_x \in [8, 33]$ keV}
		\label{fig:subfig_5pl_ns2}
	   \end{subfigure}
\vfill
\begin{subfigure}{0.49\linewidth}
		 \includegraphics[width=\textwidth]{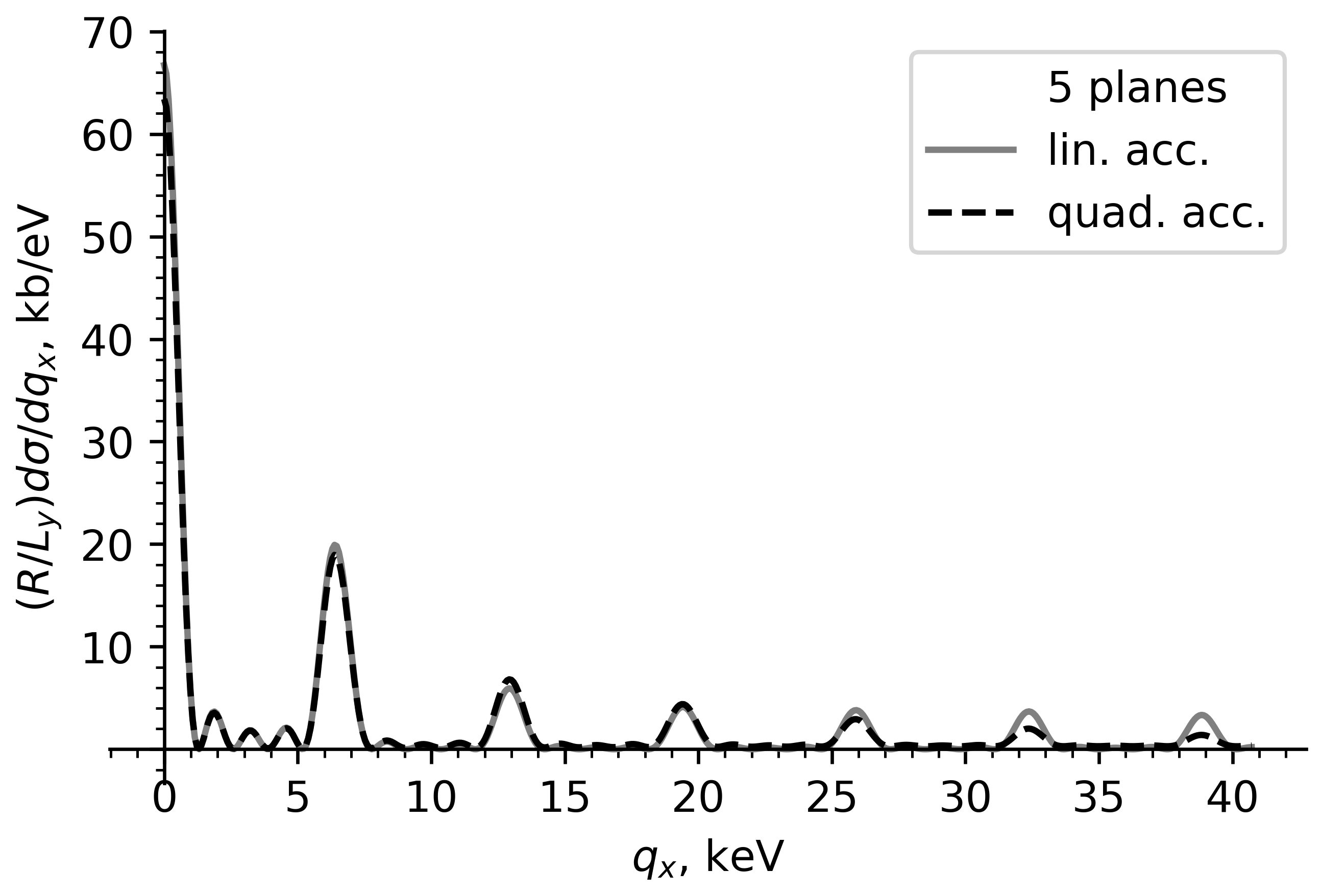}
		 \caption{accuracy up to linear and quadratic potential terms, $q_x \in [0, 40.8]$ keV}
		 \label{fig:subfig_5pl_lq0}
	      \end{subfigure}
	     \begin{subfigure}{0.49\linewidth}
		 \includegraphics[width=\textwidth]{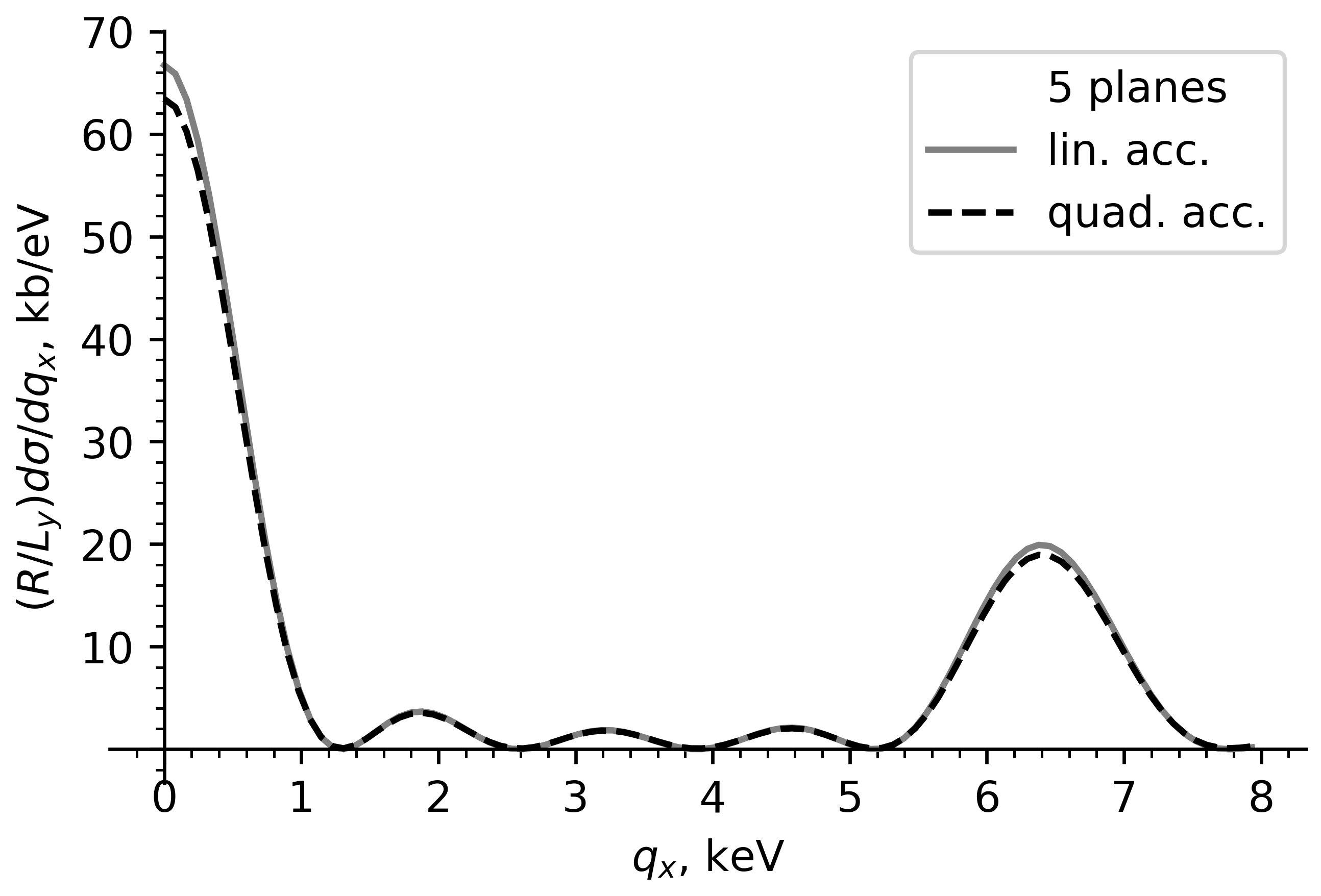}
		 \caption{accuracy up to linear and quadratic potential terms, $q_x \in [0, 8]$ keV}
		 \label{fig:subfig_5pl_lq1}
	      \end{subfigure}

\vfill

 \begin{subfigure}{0.49\linewidth}
		 \includegraphics[width=\textwidth]{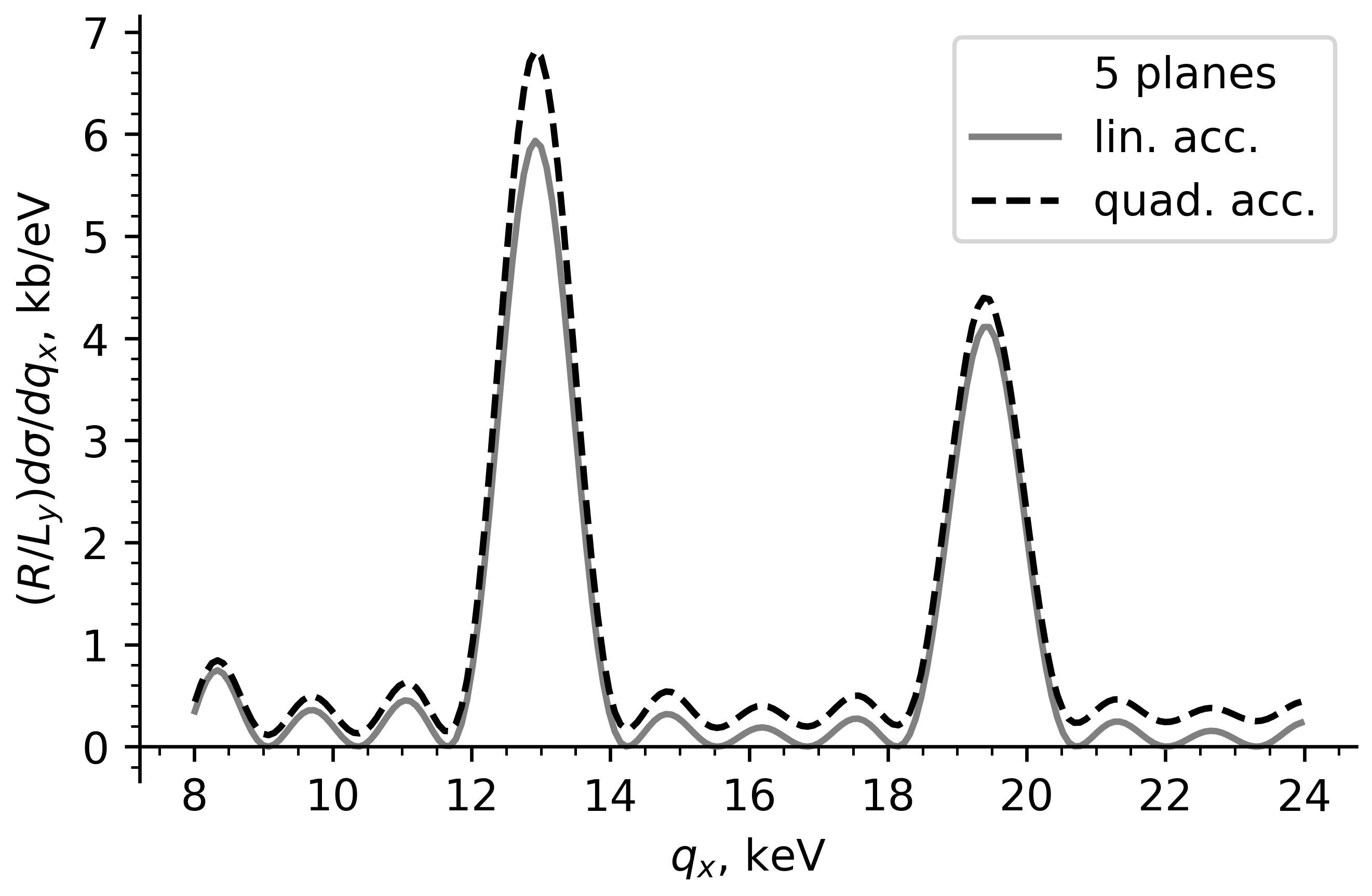}
		 \caption{accuracy up to linear and quadratic potential terms, $q_x \in [8, 24]$ keV}
		 \label{fig:subfig_5pl_lq2}
	      \end{subfigure}
\begin{subfigure}{0.49\linewidth}
		 \includegraphics[width=\textwidth]{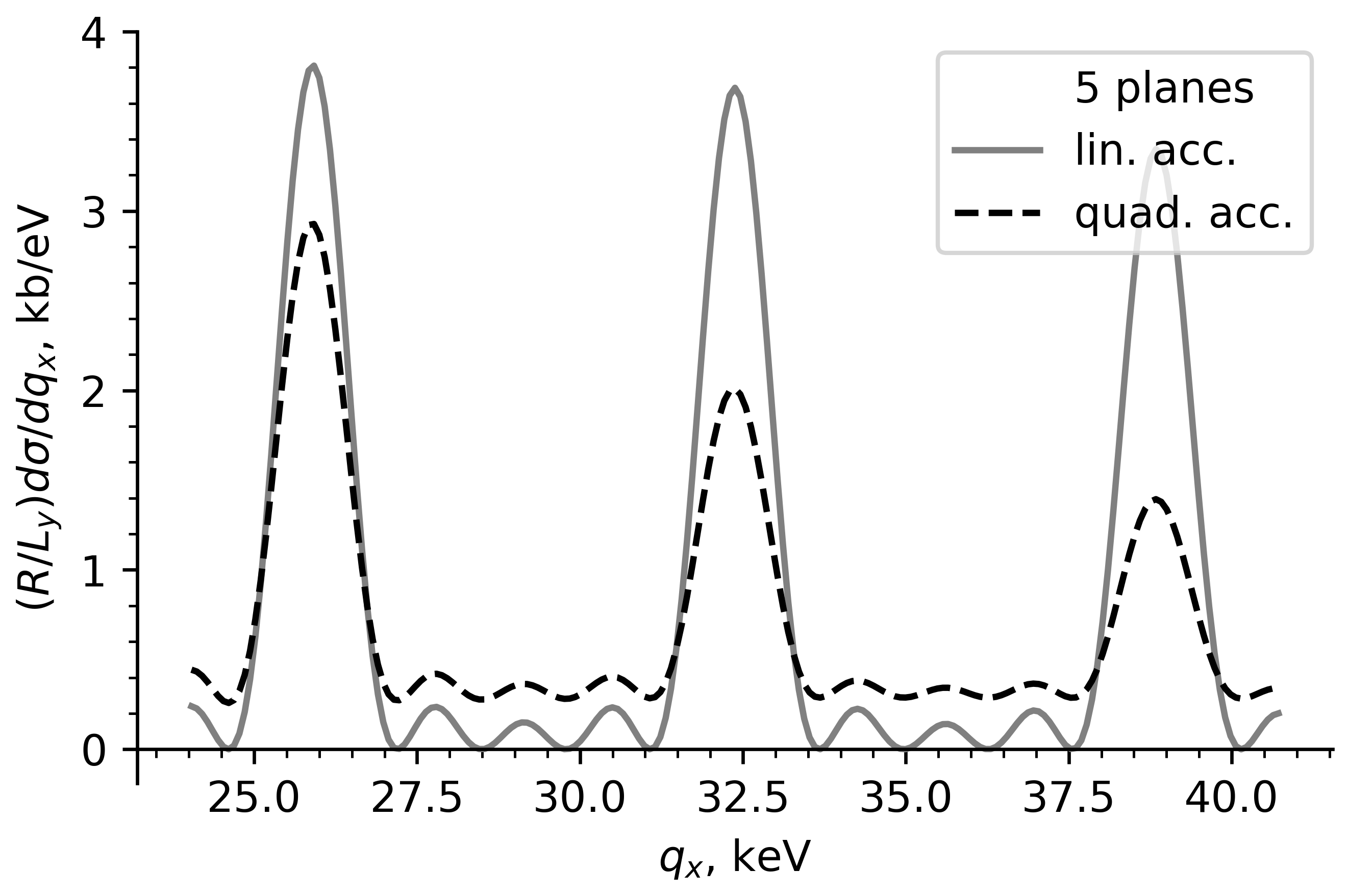}
		 \caption{accuracy up to linear and quadratic potential terms, $q_x \in [24, 40.8]$ keV}
		 \label{fig:subfig_5pl_lq2}
	      \end{subfigure}
\caption{The differential cross section of fast charged particles scattering on 5 atomic planes}
 \label{fig_pl5}
\end{figure}

Results of calculations for 1-5 atomic planes in the target are presented in Figs. \ref{fig_pl1}-\ref{fig_pl5}. As we see from Figs. \ref{fig_pl2}-\ref{fig_pl5} results of calculation with formulas \eqref{eq12_1} and \eqref{eq37} agree well. They are presented in figures with labels "numerical integration and structure factors approaches", where "numerical integration" labels calculations with formula  \eqref{eq12_1} and "structure factors approach" -- with \eqref{eq37}. 

Let us compare obtained differential cross sections with corresponding results in \cite{Born} in the Born approximation. In the Born approximation, the differential cross section of scattering on $N_x$ planes with uniform distribution of atoms could be written in terms of current paper as
\begin{eqnarray}\label{eq38}
\frac{d \sigma_{Born}^{(N_x)}}{d q_x}=N_p \left\{N_x \frac{d \sigma_{Born}^{atom}}{dq_x} + 2\pi n_{yz} L_z D_{N_x} e^{-q_x^2 \langle u_x^2 \rangle} \frac{d^2 \sigma_{Born}^{atom}}{dq_x dq_y}\Big|_{q_y=0} \right\},
\end{eqnarray}
where $\frac{d^2 \sigma_{Born}^{atom}}{dq_x dq_y}$ is the differential cross section of scattering on one atom in the Born approximation, $\frac{d \sigma_{Born}^{atom}}{dq_x}=\int_{-\infty}^{\infty} dq_y \frac{d^2 \sigma_{Born}^{atom}}{dq_x dq_y}$, $\langle u_x^2 \rangle$ is average square value of thermal vibrations of atoms in the lattice. In the present paper we consider the target without accounting for thermal vibrations, meaning $\langle u_x^2 \rangle =0$.  
If we set $N_x=1$, then $D_{N_x}=1$ and the differential cross section for scattering on a single plane in the Born approximation is 
\begin{eqnarray}\label{eq39}
\frac{d \sigma_{Born}^{(1)}}{d q_x}=\frac{d \sigma_{B, coh}^{(1)}}{dq_x}+\frac{d \sigma_{B, incoh}^{(1)}}{dq_x},
\end{eqnarray}
where $\frac{d \sigma_{B, coh}^{(1)}}{dq_x}=2\pi n_{yz} L_z e^{-q_x^2 \langle u_x^2 \rangle} \frac{d^2 \sigma_{Born}^{atom}}{dq_x dq_y}\Big|_{q_y=0}$ and $\frac{d \sigma_{B, incoh}^{(1)}}{dq_x}=N_p \frac{d \sigma_{Born}^{atom}}{dq_x}$ are, respectively, coherent and incoherent differential cross sections of scattering on a single plane with uniformly distributed atoms in the Born approximation. In terms of these values, formula \eqref{eq38} is
\begin{eqnarray}\label{eq40}
\frac{d \sigma_{Born}^{(N_x)}}{d q_x}= N_x \frac{d \sigma_{B, incoh}^{(1)}}{dq_x} + D_{N_x} \frac{d \sigma_{B, coh}^{(1)}}{dq_x}.
\end{eqnarray}
If we compare \eqref{eq37} and \eqref{eq40}, we see that these formulas for cross sections in the eikonal and Born approximation have similar structure. In both formulas there is a part depending only on number of planes in the target and a part depending on the structure factor. Using analogy, we can rewrite \eqref{eq37} as
\begin{eqnarray}\label{eq41}
\frac{d \sigma^{(N_x)}}{d q_x}= N_x \frac{d \sigma_{eik, incoh}^{(1)}}{dq_x} + D_{N_x} \frac{d \sigma_{eik, coh}^{(1)}}{dq_x},
\end{eqnarray}
where $\frac{d \sigma_{eik, incoh}^{(1)}}{dq_x}=\frac{L_y}{2\pi} \left( -I_1^{(1)} - I_2^{(1)} +I_3^{(1)} \right)$ and $\frac{d \sigma_{eik, coh}^{(1)}}{dq_x}=\frac{L_y}{2\pi} I_0^{(1)}$. Then we can label $\frac{d\sigma_{incoh}^{(N_x)}}{dq_x}=N_x \frac{d \sigma_{eik, incoh}^{(1)}}{dq_x}$ and $\frac{d \sigma_{coh}^{(N_x)}}{dq_x}=D_{N_x} \frac{d \sigma_{eik, coh}^{(1)}}{dq_x}$.
This analogy between eikonal and Born approximations makes sense for scattering on the set of atomic planes in the following way. There is a relatively small for small transferred momenta part of the cross section (Fig. \ref{fig:subfig_1pl_ci}) which contribution to the total differential cross section is proportional to number of planes. Also, there is a part depending on the structure of the target. For a target with ordered planes this term leads to significant increase of the differential scattering cross section for some values of transferred momenta. For $N_x \gg 1$: $D_{N_x} \sim N_x \sum_{g_j}\delta({q_x-\tilde{g}_j})$, where $\tilde{g}_j=\frac{2\pi j}{a}$ are reciprocal lattice vectors and $j=0, \pm 1, \pm 2, ...$. This property indicates that the closer the transferred momenta to any value of $\tilde{g}_j$, the greater increment the differential scattering cross section receives. 
It is worth noting that for large transferred momenta the differential scattering cross section in the eikonal approximation is majorly defined by the incoherent scattering cross section (Fig. \ref{fig:subfig_1pl_ci}). The same property is also correct for the Born approximation since $\frac{d \sigma_{B, coh}^{(1)}}{dq_x} \sim e^{-q_x^2 \langle u_x^2 \rangle}$, meaning that results in the Born and eikonal approximation also agree in this aspect.    

Also we note that we stated that there was no separation of the coherent and incoherent cross sections in the differential scattering cross section in the eikonal approximation for the case of scattering on a single plane \cite{Eik}. The eikonal approximation may be interpreted in the following way: the distribution of atoms in the target along $z$-axis (the direction of initial incident particle momenta) does not make any contribution to the result while applicability conditions are met. The only spatial structure that influences the scattering is what we could see in the perpendicular plane to the direction of the initial particle momenta (in the $(x,y)$-plane). If there are spatially separated projections of substructures on the $(x,y)$-plane then we can apply an approach described in the current paper and obtain coherent and incoherent scattering cross sections separately. For a set of parallel atomic planes with the spacing between them like between $(100)$ Si planes, the atomic planes are spatially separated. But if we consider atoms in the plane with uniform distribution of atoms we will see approximately solid line segment in the $(x,y)$-plane, meaning there is no sufficient separation between atoms. So, the approach considering isolated substructures cannot be used to describe scattering on a single plane with uniformly distributed atoms. That is why we did not see the separation of the coherent and incoherent differential cross sections for scattering on a single plane in the eikonal approximation in \cite{Eik}. 

Also, the results of this paper indicate that even in the model without thermal vibrations, the incoherent scattering cross section occurs.

Also we see from  Figs. \ref{fig_pl1}-\ref{fig_pl5} with label "accuracy up to linear and quadratic potential terms" that the "linear" approximation (considering only linear on potential terms) which corresponds to the continuous potential approximation describes main pattern given by $D_{N_x}$ factor. Also "linear" approximation gives quite close differential cross sections to those calculated with quadratic potential terms for small enough transferred momenta. When transferred momenta become large enough the differential cross section is mainly defined by the incoherent cross section. In this case "quadratic" approximation becomes important.   

We also note that obtained differential scattering cross sections have oscillatory patterns. This oscillations are also a manifestation of quantum rainbow effect. As known, quantum rainbow appears in cases when dependence of trasferred momentum (or deflection angle) on the impact parameter from the quasi-classical point of view ($q_x=-\partial_x \chi_0^{(N_x)}(x)$) has extrema. Existence of extrema is equivalent to the inverse function $x(q_x)$ being multivalued, meaning some transferred momenta can be a result of particle incidence with different impact parameters. In our case $\partial_x \chi_0^{(N_x)}(x)$ is periodic as well as $\chi_0^{(N_x)}$-function, meaning $x(q_x)$ in this case is multivalued. That leads to quantum rainbow effect which manifests in the differential scattering cross section oscillations.

This page is followed by Figs. \ref{fig_pl4}, \ref{fig_pl5}.
\FloatBarrier
\begingroup

\section*{Acknowledgements}
\sloppy
{Author thanks God and His Blessed Mother for saving us and our Ukraine. The work was partially supported by the National Academy of Sciences of Ukraine (project 0124U002155). The author is deeply thankful to her late research supervisor, an academic of NAS of Ukraine N.F.~Shul'ga, who introduced this problem to her and gave a lot of necessary knowledge. Author also acknowledges the fruitful discussions with S.P. Fomin, I.V. Kyryllin, \linebreak S.V.~Trofymenko.  
\par}

\endgroup

\end{document}